\documentclass[a4paper,11pt]{article}
\pdfoutput=1 
\usepackage{jheppub} 

\def\CA{{\cal A}}

\def\CC{{\cal C}}

\def\CH{{\cal H}}

\def\CN{{\cal N}}
\def\CO{{\cal O}}

\def\CS{{\cal S}}

\newcommand{\vev}[1]{ \left\langle {#1} \right\rangle }

\def\diag{\mathop{\rm diag}\nolimits}

\def\tr{\mathop{\rm tr}}

\def\beq#1\eeq{\begin{align}#1\end{align}}

\title{The moduli space of vacua of $\CN=2$ class $\CS$ theories}

\author[]{Dan Xie}
\author[]{and Kazuya Yonekura}
\affiliation[]{School of Natural Sciences, Institute for Advanced Study, 1 Einstein Drive, \\
Princeton, NJ 08540, USA}

\abstract{We develop a systematic method to describe the moduli space of vacua of four dimensional $\mathcal{N}=2$ class ${\cal S}$ theories including 
Coulomb branch, Higgs branch and mixed branches.
In particular, we determine the Higgs and mixed branch roots, 
and the dimensions of the Coulomb and Higgs components of mixed branches. 
They are derived by using generalized Hitchin's equations obtained from twisted compactification of 5d maximal Super-Yang-Mills,
with local degrees of freedom at punctures given by (nilpotent) orbits. 
The crucial thing is the holomorphic factorization of the Seiberg-Witten curve  
and reduction of singularity at punctures.
We illustrate our method by many examples including $\CN=2$ SQCD, $T_N$ theory and Argyres-Douglas theories.
}

\begin{document} 
\maketitle
\flushbottom

\section{Introduction}

The structure of moduli space of vacua plays a crucial role in studying the dynamics of
 supersymmetric field theory. Sometimes one
can find exact solutions by analyzing various properties of the moduli space, such as the singularity structure, asymptotical behavior, etc; and the typical example is the
Seiberg-Witten solutions of  $\mathcal{N}=2$ theory~\cite{Seiberg:1994rs,Seiberg:1994aj}. 

\begin{center}
\begin{figure}[htbp]
\small
\centering
\includegraphics[width=8cm]{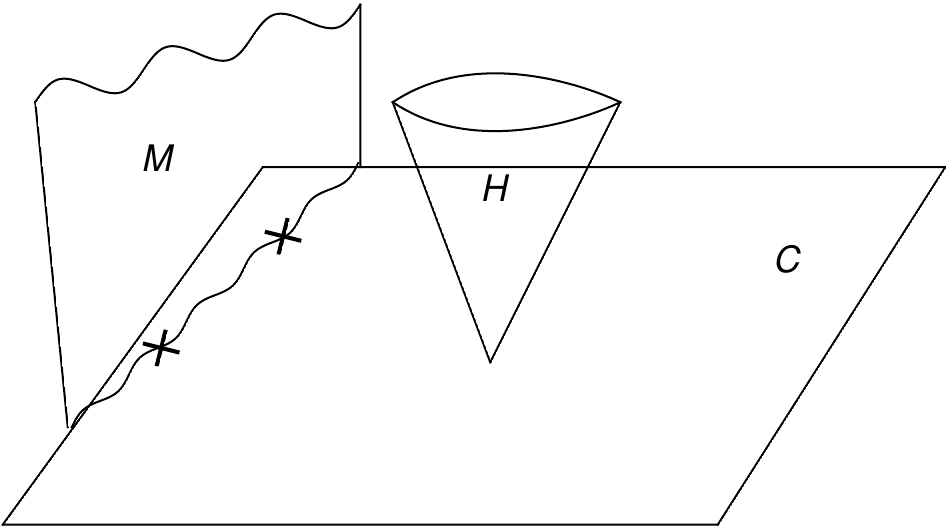}
\caption{The vacuum structure of $\CN=2$ theory: C: Coulomb branch; H: Higgs branch; M: Mixed branch.}
\label{intro}
\end{figure}
\end{center}

The full moduli space of four dimensional $\mathcal{N}=2$ theory has the general structure
\beq
\bigcup_{\alpha } \CC_\alpha \times \CH_\alpha, \label{eq:intromixed}
\eeq
where $\alpha$ labels the components of branches, and
 $\CC_\alpha$ are parametrized by Coulomb moduli fields and $\CH_\alpha$ are parametrized by Higgs moduli fields.
The $\CC_\alpha$ are special Kahler manifolds and the $\CH_\alpha$ are hyperkahler manifolds, and their metrics do not mix due to $\CN=2$
supersymmetry~\cite{deWit:1984px,Argyres:1996eh}.
There is usually a pure Coulomb branch, which we simply denote as $\CC$; 
sometimes, there is also a pure Higgs branch $\CH$ which touches with the Coulomb branch at a single point (such as SCFT points). 
Mixed branches are emanating from special loci of the Coulomb branch as shown in figure~\ref{intro}, 
and the corresponding submanifolds on the Coulomb branch are called 
the Higgs or mixed branch roots \cite{Argyres:1996eh}. 

The  Coulomb branch of $\mathcal{N}=2$ theory has been studied extensively, and the Seiberg-Witten solution can often be found easily
by using the brane construction \cite{Witten:1997sc, Landsteiner:1997vd, Brandhuber:1997cc}, or using the relation to integrable systems \cite{Donagi:1995cf,Martinec:1995by}. 
The pure Higgs branches are much less studied, 
see e.g.,~\cite{Argyres:1996eh,Argyres:1996hc,Gaiotto:2008nz,Gaiotto:2011xs,Argyres:2012fu,Maruyoshi:2013hja,Beem:2013sza} for some examples. 
The full structure of moduli space is known for very few examples, i.e. the full vacua structure of 
$\mathcal{N}=2$ SQCD has been determined in \cite{Argyres:1996eh,Argyres:1996hc}  
using the non-renormalization theorem and exact solution on the Coulomb branch. See also the discussion using
type IIA brane construction \cite{Witten:1997sc,Hori:1997ab,Nakatsu:1997jw,Nakatsu:1997bt}.

We are going to study the full moduli space of $\CN=2$ theories obtained by twisted compactification of 6d $\CN=(2,0)$ $A_{N-1}$ theory on a Riemann surface,
called class $\CS$ theories~\cite{Gaiotto:2009we,Gaiotto:2009hg,Chacaltana:2010ks,Chacaltana:2012zy,Xie:2012hs}\footnote{By class $\CS$ theories, we mean four dimensional theories which can be derived 
from compactifying 6d theory on a Riemann surface with all kinds of defects, such as regular defects, irregular defects, etc. This class of theories include many
well studied theories such as SQCD, Argyres-Douglas theory \cite{Argyres:1995jj}, Maldacena-Nunez theory \cite{Maldacena:2000mw}, etc.}.
Instead of using field theory method, we are going to use the geometric method 
developed in \cite{Xie:2013gma,Xie:2013rsa,Yonekura:2013mya} (see also \cite{Bonelli:2013pva}) to determine various branches. 

Let's first recall some of the basic features of  M5 brane construction which leads to this class of theories. 
There are five scalars $\varphi_I~(I=1,2,3,4,5)$ in the adjoint representation of $A_{N-1}=SU(N)$ which
describe the transverse fluctuation of M5 branes. 
Four dimensional theory is derived by twisted compactification of the 6d theory on a Riemann surface $C$, 
and one get a Higgs field $\Phi$  in the canonical (or cotangent)
bundle of $C $ which describes the Coulomb branch deformations. There are other real scalars $\vec{\varphi}=(\varphi_1,\varphi_2,\varphi_3)$  
which are in the trivial bundle of $C$ and describe the Higgs branch deformations.  Generically, the Coulomb branch and Higgs branch
are determined as follows:
\begin{itemize}
\item The Coulomb branch is described by turning on only the deformations in $\Phi$, and the Seiberg-Witten curve is then given by the spectral curve of $\Phi$:
\begin{equation}
0=\det(xI_N-\Phi)= x^N+\sum_{i=2}^N \phi_i(z) x^{N-i},
\end{equation}
here $\phi_i$ are degree $i$ holomorphic differential on Riemann surface.
\item The pure Higgs branch is described by turning on only $\vec{\varphi}$ deformations  (we can get another scalar from $B_{\mu\nu}$ field on M5 brane \cite{Hori:1997ab}.).
\item Mixed branches are described by turning on both $\vec{\varphi}$ and $\Phi$.
\end{itemize}

Let's discuss the mixed branch in more detail. The deformation in $\vec{\varphi}$ is simple:
  they can be diagonalized with constant eigenvalues :
\beq
\vec{\varphi}=\diag(\vec{a}^{(1)} I_{n_1},\cdots,\vec{a}^{(k)}I_{n_k} ),
\eeq
where $n_s$ are integers such that $\sum_{s=1}^k n_s=N$, $I_{n_s}$ are $n_s \times n_s$ the unit matrices, and $\vec{a}^{(s)}$ are constants.
Now the crucial thing is that brane dynamics requires the commuting condition $[\Phi, \vec{\varphi}]=0$, which
impies that the Seiberg-Witten curve has to be factorized as:
\beq
\det (x I_N-\Phi) =x^N+\sum_{i=2}^N \phi_i x^{N-i}=0
\to \prod_{s=1}^k \left(x^{n_s}+\sum_{i=1}^{n_s} \phi_{s,i}(z)x^{n_s-i} \right)=0 \label{eq:introfacto},
\eeq
here $\phi_{s,i}(z)$ are various holomorphic differentials, and $n_s$ are the same as above.

\begin{center}
\begin{figure}[htbp]
\small
\centering
\includegraphics[width=8cm]{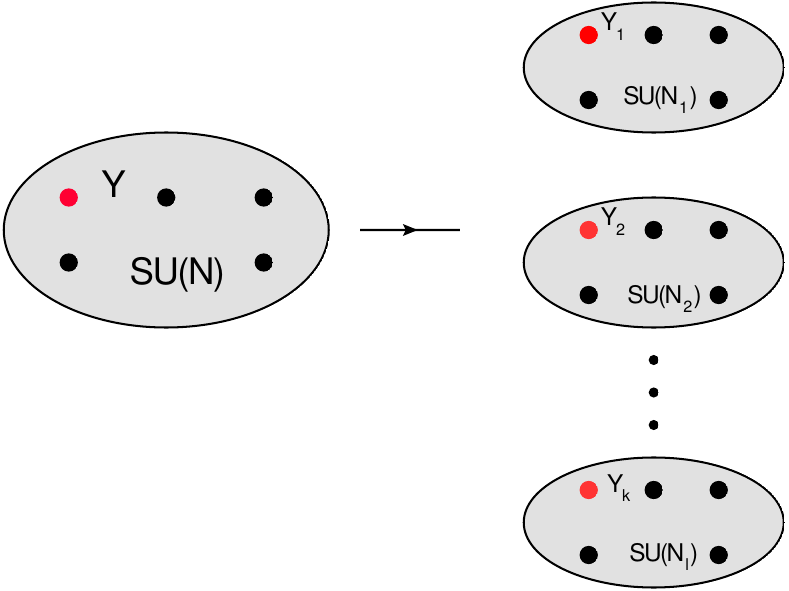}
\caption{The Higgs and mixed branch roots can be described as: the $SU(N)$ Hitchin system is decomposed into $SU(n_1)\times SU(n_2) \times\ldots\times SU(n_k) \times U(1)^{k-1}$ Hitchin systems, and
the regular punctures of $SU(N)$ are decomposed into sums of regular punctures of lower rank Hitchin systems.}
\label{intro1}
\end{figure}
\end{center}

There are more variations due to punctures on the Riemann surface $C$. 
Regular punctures are classified by Young diagrams with total number of boxes $N$, and one can define an ordering between two Young diagrams.
In the mixed branch, a Young diagram $Y^D$ of $SU(N)$ theory is split into a collection of Young diagrams $Y^D_{s}$ of $SU(n_s)$ theory, 
see figure~\ref{intro1}. (The superscript $D$ will be explained later in this paper.) One can form 
a new Young diagram $Y'^{D}$ with $N$ boxes by assembling $Y^D_{s}$ together, and the constraint on $Y'^{D}$ is that 
\begin{equation}
Y'^{D}\leq Y^D,
\end{equation}
where we used the ordering between two Young diagrams.
There is a mixed branch corresponding to each collection $Y^D_s$.

Unlike regular singularity, the constraint  from irregular singularities is more rigid, as the boundary condition of $\Phi$ is fixed at the puncture and sometimes
the curve can not be factorized in an arbitrary way. The factorization 
pattern of Seiberg-Witten curve of equation \eqref{eq:introfacto} has to be consistent with the boundary condition on the irregular puncture.

This paper is organized as follows. In section~\ref{sec:2}, we give a general algorithm of how to find mixed branch roots, and how to count the dimensions of the Coulomb and Higgs factors.
In section~\ref{sec:3} we derive those rules following the strategy outlined above.
In section~\ref{sec:4} and appendix~\ref{sec:appA}, we reproduce the results of $\CN=2$ SQCD by our method.
In section~\ref{sec:5}, theories defined using regular singularities are discussed.
In section~\ref{sec:6}, we study Argyres-Douglas theories. 
Finally in section~\ref{sec:7}, we give a short discussion.

\section{General rules for finding mixed branches}\label{sec:2}
In general, the moduli space of $\CN=2$ supersymmetric field theory has the general form
\beq
\bigcup_{\alpha } \CC_\alpha \times \CH_\alpha,
\eeq
where $\alpha$ runs over all the components of the moduli space, and $\CC_\alpha$ and $\CH_\alpha$ are Coulomb factor and Higgs factor, respectively.
In other words, $\CC_\alpha$ is parametrized by vevs of vector multiplets, and $\CH_\alpha$ is parametrized by vevs of hypermultiplets.
For Lagrangian theory, the Higgs branch can be found using classical equations of motion due to non-renormalization theorem \cite{Argyres:1996eh}.
Coulomb branch is much more difficult to determine because one needs to take into account various quantum corrections, and the result 
can be elegantly summarized using Seiberg-Witten curve.

In this section we would like to state the rules of how to find the roots of these mixed branches $\CC_\alpha \times \CH_\alpha$
and give the formulas for the dimensions of the Coulomb factor $\CC_\alpha$ and the Higgs factor $\CH_\alpha$.
Derivations of the rules are given in the next section~\ref{sec:3}.
One of the branches is the pure Coulomb branch $\CC \times \{0\}$, which we denote as just $\CC$.
Then, the root of the branch $\CC_\alpha \times \CH_\alpha$ is the intersection of it with $\CC$.

\subsection{Pure Coulomb branch of class $\CS$ theories}
The Seiberg-Witten curve describing the Coulomb branch $\CC$ is given by the general form~\cite{Gaiotto:2009we,Gaiotto:2009hg,Xie:2012hs},
\beq
0=x^N+\sum_{i=2}^N \phi_i(z) x^{N-i},\label{eq:SWGcurve}
\eeq
where $z$ is the coordinate of the base Riemann surface $C$, and $x$ is the coordinate of the fiber of the canonical (or cotangent) bundle $K=T^*C$.
The $\phi_i(z)$ is a section of the line bundle $K^{\otimes i}$, i.e., it is an $i$-th differential on $C$. 
Neglecting punctures, its contribution to the dimension of Coulomb moduli can be found using the 
Riemann-Roch theorem,
\begin{equation}
\dim(\phi_i)=\dim H^0(C, K^{\otimes i})=(2i-1)(g-1),
\end{equation}
where we used the vanishing theorem $H^0(C,K^{1-i})=0$ for $i>1$.

On the Riemann surface $C$, we can also add various regular and irregular punctures.
Let us recall some properties of them. We consider local properties near one puncture, and we take the coordinate so that the puncture is at $z=0$ for regular singularity.

\paragraph{Regular singularities.}
Regular (or tame) singularities~\cite{Gaiotto:2009we} are classified by a partition of $N$, $[m_1,\cdots, m_\ell]$, which satisfies $N=m_1+\cdots+m_\ell$.
Without loss of generality we take $m_1 \geq m_2 \cdots \geq m_\ell$. This partition can be represented by a 
Young diagram $Y$ such that the $a$-th column has $m_a$ boxes (i.e., its hight is $m_a$).
In our convention, a full puncture is represented by $[1,1,\ldots, 1]$, and a simple puncture is represented by $[N-1,1]$.

\begin{center}
\begin{figure}[t]
\small
\centering
\includegraphics[width=12cm]{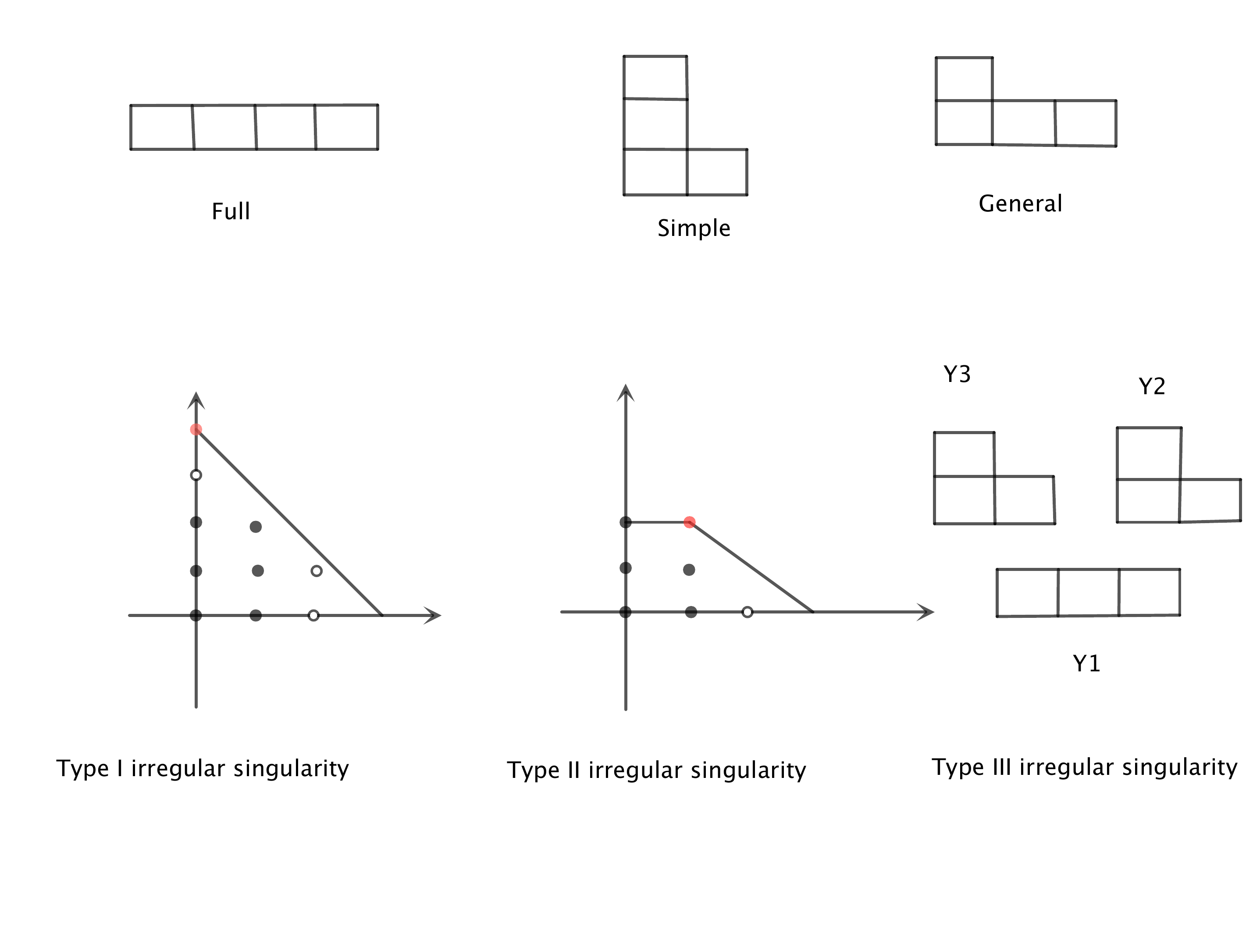}
\caption{Upper: regular puncture are represented by Young diagrams. Bottom: Two types of irregular singularities are represented 
by Newton polygon, and a third type is represented by a collection of Young diagrams.}
\label{irregular}
\end{figure}
\end{center}

The order of pole $p_i$ at the regular singularity of the differential $\phi_i(z)$ can be easily found from the Young diagram\footnote{
Unless otherwise stated, we take mass parameters of regular singularities to be zero.}
\beq
\phi_i(z) &=\CO(z^{-p_i}),~~~~~
p_i =i-s_i, \label{eq:degofsing}
\eeq
where $s_i$ is the height of the $i$th box of the Young diagram. The counting is from left to right and from bottom to top. 
See figure~\ref{irregular} for examples of Young diagrams.

The local contribution of a regular puncture  to the Coulomb branch dimension is 
\beq
d(Y^D)=\sum_{i=2}^N p_i=\frac{1}{2} \left(N^2-\sum_{a=1}^{\ell} m_a^2 \right),\label{eq:orbitdim}
\eeq
where $Y^D$ is the Young diagram obtained by transposing $Y$, i.e., the rows of $Y^D$ are columns of $Y$.
The reason for using $Y^D$ in \eqref{eq:orbitdim} will become clearer in the next section.

For later use, let us define partial ordering of two Young diagrams. We define
\begin{equation}
Y^D \geq Y'^D \Longleftrightarrow p_i\geq p^{'}_i  \Longleftrightarrow \sum_{b=1}^{a} \tilde{m}_a \geq \sum_{b=1}^{a} \tilde{m}'_a. 
\label{eq:partitionorder}
\end{equation}
where $Y^D=[\tilde{m}_a]$. The second equivalence can be checked by using \eqref{eq:degofsing}.
Notice that this is only a partial ordering: it is possible that neither $Y^D \geq Y'^D$ nor $Y^D \leq Y'^D$.

\paragraph{Irregular singularities.}

We can also have irregular (or wild) singularities, which are needed for asymptotically free theories and 
Argyres-Douglas theories. The irregular singularities which are relevant for SQCD will be discussed in detail in appendix~\ref{sec:appA}.
The irregular singularities which are relevant for Argyres-Douglas theories can also be  summarized by a Newton polygon, 
see figure \ref{irregular}, and the local contribution to the Coulomb moduli can also be found from Newton polygon, for details, see  \cite{Xie:2012hs}.
The leading order eigenvalues of $x$ can be read from the slop of the Newtwon polygon:
\begin{align}
&\text{Type I}: x\sim {1\over z^{2+{k\over N}}}(1, \omega_N, \ldots, \omega_N^{N-1} )     \nonumber\\
&\text{Type II}: x\sim {1\over z^{2+{k\over N-1}}}(0,1, \omega_{N-1}, \ldots, \omega_{N-1}^{N-2})     \nonumber \\
&\text{Type III}: x\sim {A_n\over z^n}+{A_{n-1}\over z^{n-1}}+\ldots+{A_1\over z}
\end{align}
here $k$ is the height of the marked point in the Newton polygon of figure \ref{irregular}, $\omega_\ell=\exp(2 \pi i/\ell)$,
and the eigenvalue degeneracy of $A_i$ is determined by the Young diagrams $Y_i$.

The major feature of irregular singularity is that the leading order behavior is fixed as the UV data, and we can not change it by tuning 
the Coulomb branch moduli.
 For regular singularities, one can actually tune the Coulomb branch moduli to change the singularity.

\subsection{Mixed branches}\label{sec:mix}
Now we can state our general rules for finding the roots of mixed branches $\CC_\alpha \times \CH_\alpha$.

\paragraph{Step 1.}
The first step is to choose a partition of $N$: $X=[n_1,n_2,\cdots, n_k]$. 
Then we take the curve \eqref{eq:SWGcurve} to be of factorized form
\beq
x^N+\sum_{i=2}^N \phi_i(z) x^{N-i}=0 \to
\prod_{s=1}^k \left(x^{n_s}+\sum_{i=1}^{n_s} \phi_{s,i}(z)x^{n_s-i} \right)=0, \label{eq:facto}
\eeq
where $\phi_{s,i}$ are holomorphic differentials on the Riemann surface, and we also have
 $\sum_s \phi_{s,1}=0$ due to the traceless condition.  
 
 The existence of  type I and type II irregular singularities would constrain the 
possible factorization $X$:
\begin{itemize}
\item Type I irregular singularity: 
define $d$ as the maximal common divisor of $(k,N)$. 
Then the maximal partition $X$ is 
\begin{equation}
X_{\rm max}=[{N\over d},{N\over d},\ldots, {N\over d}],
\end{equation}
and other possible factorization is derived by combining the columns of  $X$.
\item Type II irregular singularity: 
define $d$ as the maximal common divisor of $(k,N-1)$, then the maximal partition $X$ is
\begin{equation}
X_{\rm max}=[{{N-1}\over d},{N-1\over d},\ldots, {N-1\over d},1]
\end{equation}
and other partitions are found by combining  columns expect the last column with height one. 
\end{itemize}
When there are more than one irregular singularity, $X$ is constrained to be compatible with all irregular singularities. 

\paragraph{Step 2.}
The second step is to take into account local punctures.  After the above global factorization, the  $SU(N)$ theory is 
split into a sum of  $SU(n_s)$ (and $U(1)$) theories. For each regular puncture, 
a dual Young diagram $Y^D$ of the $SU(N)$ theory is split into a sum of diagrams 
$Y_s^D$ for $SU(n_s)$ theories. The important point is that these $Y_s^D$ are not uniquely determined from
$Y^D$. A choice of the collection $Y_s^D$ corresponds to the choice of a mixed branch we consider.

One can form a Young diagram of $N$ boxes $Y'^D$ by combining $Y_s^D$. 
The columns of $Y'^D$ are taken to be the columns of $Y_s^D$.
Then the only constraint on $Y_s^D$ is
\begin{equation}
Y'^D\leq Y^D \label{constraint}.
\end{equation}

\paragraph{Step 3.}
We consider every possible set of $X$ and $Y^D_s$, and find the curve consistent with the above conditions.
If \eqref{eq:facto} cannot be further factorized in a generic curve, this is the root of a branch $\CC_\alpha \times \CH_\alpha$,
where the label $\alpha$ is specified as $\alpha=\{X, Y^D_s\}$.
If \eqref{eq:facto} is factorized in a generic curve, this is just a boundary of some other mixed branch and we discard this case.

Now let us give the formulas for the dimensions of the branches. 
\paragraph{Coulomb factor.} Once the factorization and local data are given, 
the dimension of the Coulomb factor $\CC_\alpha$ can be easily found from 
the factorized Seiberg-Witten curve: each factor is the spectral curve of a lower rank $SU(n_s)$ or $U(1)$ Hitchin system. 
Each $U(1)$ Hitchin system contributes dimension $g$ to the Coulomb branch dimension.

Here we give an explicit formula for the Coulomb branch factor if there are only regular punctures. First, let us define local contribution of a puncture $p$ as
\beq
\dim_{\mathbb C} \CC_\alpha(p) \equiv \sum_{s=1}^k d(Y^D_{s}) \label{eq:localCdim},
\eeq
where $d(Y_s^D)$ is defined as in \eqref{eq:orbitdim}.
Then, the total dimension is given by 
\beq
\dim_{\mathbb C} \CC_\alpha=(k-1)g+\sum_{s=1}^k (n_s^2-1)(g-1)+ \sum_p \dim_{\mathbb C} \CC_\alpha(p),\label{eq:coulombdim}
\eeq
where $k$ is the number of elements in the partition $X=[n_1,\cdots,n_k]$. The local contribution of the irregular singularity can 
be found using the method discussed in \cite{Xie:2012hs}.

\paragraph{Higgs factor.}
The dimension of the Higgs factor $\CH_\alpha$ has several contributions. There is global contribution due to the factorization 
of the Seiberg-Witten curve. The quaternionic dimension of this part is simply $k-1$.

There is also contribution from local punctures, and an explicit form of the regular puncture contribution can be easily written down.
Defining $Y'$ as the dual Young diagram of $Y'^D$, the local Higgs branch dimension is given by
\begin{equation}
\dim_{\mathbb H} \CH_\alpha(p)=d(Y')-d(Y). \label{eq:localHdim}
\end{equation}
This local contribution is always non-negative due to the constraint on $Y^{'}$ given by \eqref{constraint}, which is actually equivalent to $Y' \geq Y$. 
There is no local contribution to Higgs branch from type I and type II irregular singularity. 
The total dimension of the Higgs branch is given by (assuming there is only regular puncture):
\beq
\dim_{\mathbb H} \CH_\alpha=k-1+\sum_p \dim_{\mathbb H} \CH_\alpha(p). \label{eq:higgsdim}
\eeq

\section{Vacuum structure of class $\CS$ theories}\label{sec:3}

In this section, we derive the rules for finding mixed branches stated in the previous section.
The reader who is only interested in applications of the rules can skip this section.
The essential ingredients are generalized Hitchin's equations and 5d maximal Super-Yang-Mills (SYM) twisted on the Riemann surface $C$
~\cite{Xie:2013gma,Xie:2013rsa,Yonekura:2013mya},
 $T^\rho[SU(N)]$ theories of Gaiotto and Witten~\cite{Gaiotto:2008ak} placed at punctures~\cite{Chacaltana:2012zy}, and irregular singularity classified in \cite{Xie:2012hs}.

\subsection{5d maximal SYM and generalized Hitchin's equations}\label{sec:bulk}

The class $\CS$ theories are defined as the low energy limit of 6d $\CN=(2,0)$ theory compactified on a Riemann surface $C$ with 
punctures. If we further compactify it on a circle $S^1$ so that the 6d theory is placed on ${\mathbb R}^{3} \times S^1 \times C$, 
we get an $S^1$ compactification of four dimensional theory (in the small area limit of $C$). 
The same configuration can be looked at in a different way.
We first compactify the theory on $S^1$ and  get 5d maximal SYM, then we study the 5d theory on ${\mathbb R}^{3}  \times C$. 
Actually, the 5d maximal SYM might be enough to compute (BPS or protected) quantities of 
the 6d $\CN=(2,0)$ theory on $S^1$~(see e.g., \cite{Douglas:2010iu,Lambert:2010iw}).
Then, denoting the are of the Riemann surface as $\CA$, we claim (see \cite{Yonekura:2013mya} for more discussions and the scope of validity 
of the claim),\footnote{See also \cite{Gaiotto:2011xs} for finite area effects.}
\beq
&{\rm Vacuum~moduli~space~of~4d~class~}\CS {\rm ~theory~on~}{\mathbb R}^3\times S^1 \nonumber \\
=~&{\rm Vacuum~moduli~space~of~6d~class~}\CN=(2,0) {\rm ~theory~on~}{\mathbb R}^3\times S^1 \times C|_{\CA \to 0} \nonumber \\
=~&{\rm Vacuum~moduli~space~of~5d~maximal~SYM~on~}{\mathbb R}^3\times C|_{\CA \to 0}.
\eeq
Furthermore, in this relation, the Coulomb branch and Higgs branch are exchanged, 
\beq
&{\rm Coulomb~(Higgs)~branch~of~4d~class~}\CS {\rm ~theory~on~}{\mathbb R}^3\times S^1 \nonumber \\
=~&{\rm Higgs~(Coulomb)~branch~of~5d~maximal~SYM~on~}{\mathbb R}^3\times C.
\eeq

The Coulomb branch of 4d theory on $S^1$ is a Hyperkahler manifold. The Coulomb branch has a distinguished complex structure which we denote as $I$.
This complex structure does not depend on the radius of $S^1$ \cite{Seiberg:1996nz},  
so we can learn the moduli space of 4d theory using complex structure $I$ of
Coulomb branch of 4d theory on $S^1$.\footnote{This is true not only for the Coulomb branch but also for the entire vacuum moduli space,
up to the caveat discussed in section~4 of \cite{Yonekura:2013mya}.} 
Moreover, in complex structure $I$ there is a fibration structure which can be identified with the Seiberg-Witten fibration over the Coulomb moduli space.

Now we are going to describe the structure of the moduli space of vacua using the 5d theory.
The 5d maximal SYM contains gauge fields $A_M~(M=0,1,2,3,4)$ and five real adjoint scalars $\varphi_I~(I=1,2,3,4,5)$.
It must also be twisted on the Riemann surface so that half of the supersymmetry is preserved by the compactification.
Then, taking the coordinate of ${\mathbb R}^3\times C$ as $(x^\mu, z)$ where $\mu=0,1,2$ and $z$ is a complex coordinate on $C$,
the fields of the 5d SYM are given by
\beq
(\vec{\varphi}, A_\mu),~~~(\Phi,A_{\bar{z}}).
\eeq
Here, $\Phi$($=\varphi_4+i\varphi_5$) is a complex adjoint field, 
which takes values in $K \otimes {\rm ad}(E)$, where $K=T^*C$ is the canonical bundle and
${\rm ad}(E)$ is the vector bundle in the adjoint representation of the gauge group. The fact that $\Phi$ is a section of the canonical bundle
$K$ comes from the twisting of $SO(2)_R \subset SO(5)_R$ $R$-symmetry to preserve eight superchages.
The $\vec{\varphi}=(\varphi_1,\varphi_2,\varphi_3)$ are real adjoint fields in ${\rm ad}(E)$, $A_\mu$ is the gauge field on ${\mathbb R}^3$
and $A_{\bar{z}}$ is the  gauge field on $C$.
The $\vec{\varphi}$ is a triplet of the $SU(2)_R=SO(3)_R \subset SO(5)_R$ $R$-symmetry. 
The combinations $(\vec{\varphi}, A_\mu)$ and $(\Phi,A_{\bar{z}})$ are taken so that
each of them form a single multiplet of the supersymmetry. An important point is that $(\Phi,A_{\bar{z}})$ corresponds to
the Coulomb moduli fields of the 4d class $\CS$ theory, and $(\vec{\varphi}, A_\mu)$ corresponds to the Higgs moduli fields of the 4d theory.

The classical equations for supersymmetric vacua of the 5d SYM are given by generalized Hitchin's equations~\cite{Xie:2013gma}
(see also \cite{Bonelli:2013pva}),\footnote{
In the generalized Hitchin system, the fields $\Phi_1=(\varphi_1+i\varphi_2)/\sqrt{2}$, $\Phi_2=\Phi$ and $A_{\bar{z}}$, are used.
We must supplement them with $\varphi_3$ and $A_\mu$ to describe the complete moduli space of vacua.}
which in general can describe theories with four superchages.
In the case of eight supercharges (i.e., 4d $\CN=2$), they are simplified as~\cite{Yonekura:2013mya}
\beq
&F_{z\bar{z}}+[\Phi, \Phi^\dagger]=0 \label{eq:bps1} \\
&D_{\bar{z}}\Phi=0 \label{eq:bps2} \\
&D_{{z}}\vec{\varphi}=D_{\bar{z}}\vec{\varphi}=0 \label{eq:bps3} \\
&[\vec{\varphi},\Phi]=0 \label{eq:bps4} \\
&[{\varphi}_i, {\varphi}_j]=0~~(i,j=1,2,3) \label{eq:bps5}
\eeq
where $F_{z\bar{z}}=\partial_z A_{\bar{z}}-\partial_{\bar{z}} A_z+[A_z, A_{\bar{z}}]$.

The Coulomb branch of the 4d theory (or equivalently the Higgs branch of 5d theory) is described by the first two equations which are precisely
the Hitchin's equations \cite{Hitchin:1986vp,Hitchin:1987mz}. This 
branch can be determined using the classical Hitchin's equations due to non-renormalization theorem. 

The classical configuration of the Higgs branch of the 4d theory (or Coulomb branch of 5d theory) can be described by the expectation values of 
the scalar fields ${\varphi}_i$, whose 
solutions are just constant matrices, and there are various quantum corrections to the metric of the moduli space which are difficult to calculate. 
However, for our purpose of studying the mixed branch structure and counting dimensions of moduli space,
we can still use the classical equations due to the nonrenormalization theorem discussed in \cite{Yonekura:2013mya}. 
Due to the commuting condition between $\Phi$ and ${\varphi}_i$, 
we get the breaking pattern on the Hitchin's equations, and therefore determine the mixed branch roots.

\subsection{Factorization of the Seiberg-Witten curve}
Let us solve these equations in detail. From \eqref{eq:bps5},
the $\vec{\varphi}$ are simultaneously diagonalized by $SU(N)$ gauge transformations.
Let us assume that this is done. Then we get
\beq
\vec{\varphi}= \left( \begin{array}{cccc}
\vec{a}^{(1)} I_{n_1} & 0 & \ldots  &0 \\
0 & \vec{a}^{(2)} I_{n_2} & & \vdots \\
\vdots& & \ddots &  0\\
0&\ldots& 0 & \vec{a}^{(k)} I_{n_k} 
\end{array}
\right) \label{eq:constdiagvev}
\eeq
Here $N=n_1+n_2+\cdots+n_k$ defines a partition of $N$, $X=[n_s]=[n_1,\cdots,n_k]$.
The $I_{n_s}$ are $n_s \times n_s$ unit matrices, and $\vec{a}^{(s)}=({a}^{(s)}_1,{a}^{(s)}_2,{a}^{(s)}_3)$ are vectors
with $\vec{a}^{(s)} \neq \vec{a}^{(t)}$ for $s \neq t$ and $\sum_{s=1}^k n_s \vec{a}^{(s)}=0$.
These $\vec{a}^{(s)}$ are constants because of \eqref{eq:bps3}. The $SU(N)$ gauge symmetry is broken to a subgroup
$H=SU(n_1) \times \cdots \times SU(n_k) \times U(1)^{k-1}$. Furthermore, \eqref{eq:bps3} and \eqref{eq:bps4} gives
a constraint 
\beq
\Phi= \left( \begin{array}{cccc}
\Phi^{(1)} & 0 & \ldots  &0 \\
0 & \Phi^{(2)} & & \vdots \\
\vdots& & \ddots &  0\\
0&\ldots& 0 & \Phi^{(k)}
\end{array}
\right) ,~~~
A_{\bar{z}}&= \left( \begin{array}{cccc}
A_{\bar{z}}^{(1)} & 0 & \ldots  &0 \\
0 & A_{\bar{z}}^{(2)} & & \vdots \\
\vdots& & \ddots &  0\\
0&\ldots& 0 & A_{\bar{z}}^{(k)}
\end{array}
\right) \label{eq:reducedHitchinfields}
\eeq
where $\Phi^{(s)}$ and $A_{\bar{z}}^{(s)}$ are $n_s \times n_s$ matrices. Then \eqref{eq:bps1} and \eqref{eq:bps2} are equivalent to
\beq
&F^{(s)}_{z\bar{z}}+[\Phi^{(s)}, (\Phi^{(s)})^\dagger]=0,~~~D_{\bar{z}}\Phi^{(s)} \equiv \partial_{\bar{z}}\Phi^{(s)}+[A^{(s)}_{\bar{z}},\Phi^{(s)}]=0,
\label{eq:reducedHitchineq}
\eeq
for each $s=1,\cdots,k$. Therefore, each pair $(A^{(s)}_{\bar{z}},\Phi^{(s)})$ satisfies Hitchin's equations.
If the vevs of $(A^{(s)}_{\bar{z}},\Phi^{(s)})$ completely break the $SU(n_s)$ gauge symmetries, the unbroken gauge group is $U(1)^{k-1}$.
Then, taking the zero mode of $A_\mu~(\mu=0,1,2)$ on $C$ and dualizing these gauge fields to scalar fields on ${\mathbb R}^3$,
 we get dual scalars $b^{(s)}~(s=1,\cdots,k)$ with $\sum_s n_s b^{(s)}=0$.

For the moment, we neglect punctures. Then, the general structure is as follows. 
Mixed branches are labelled by the partition of $N$, $X=[n_s]$.
The Higgs factor $\CH_\alpha^{\rm bulk}$ consists of the moduli fields $(\vec{a}^{(s)}, b^{(s)})$ which contribute to the 
quaternionic dimension of $\CH_\alpha$ as $k-1$ (or real dimension $4(k-1)$). Here we added the word ``bulk'' 
to $\CH_\alpha^{\rm bulk}$ because we are neglecting punctures.
The Coulomb factor $\CC_\alpha$ consists of the solutions of Hitchin's equations with the 
gauge group $H=SU(n_1) \times \cdots \times SU(n_k) \times U(1)^{k-1}$.

If generic solutions of Hitchin's equations for $(A_{\bar{z}}^{(s)},\Phi^{(s)})$ do not break $SU(n_s)$ completely, 
the above branch is just a boundary of a more larger branch. For example, consider a simple case where
the only solution to the $SU(n_s)$ Hitchin's equations is trivial, $(A_{\bar{z}}^{(s)},\Phi^{(s)})=0$.
This can happen e.g., when the genus is $g=0$ and there are no punctures.
Then, we can smoothly go to another branch as
\beq
\vec{a}^{(s)} I_{n_s} \to \diag(\vec{a}^{(s,1)},\cdots,\vec{a}^{(s,n_s)}  )
\eeq
while satisfying \eqref{eq:bps1}-\eqref{eq:bps5}. This means that we can go to the branch in which $n_s \to 1+\cdots+1$.
More generally, if $(A_{\bar{z}}^{(s)},\Phi^{(s)})$ breaks $SU(n_s)$ to a nontrivial subgroup $H'_s \subset SU(n_s)$, 
we can smoothly turn on a vev of $\vec{\varphi}$ in the Cartan subalgebra of the unbroken group $H'_s$.
The integer $n_s$ is further partitioned to $n_s =n_{s,1}+n_{s,2}+\cdots$.
So we do not consider the original partition $[n_s]$ as a separate mixed branch.

The spectral curve of the original Hitchin system is given as 
\beq
0=\det (x I_N-\Phi)=x^N+\sum_{i=2}^N \phi_i(z) x^{N-i}. 
\eeq
Under the decomposition \eqref{eq:reducedHitchinfields}, the curve is factorized as:
\beq
\det (x I_N-\Phi) =x^N+\sum_{i=2}^N \phi_i x^{N-i}=0
\to \prod_{s=1}^k \left(x^{n_s}+\sum_{i=1}^{n_s} \phi_{s,i}(z)x^{n_s-i} \right)=0.
\eeq

\paragraph{Constraint on factorization from Irregular singularity.}

If there are irregular singularities, the factorization pattern of the SW curve is constrained because 
the irregular singularity can not factorize in an arbitrary way. The reason is that the leading order term is 
fixed and define the UV theory, so we can not tune the leading order behavior of the SW curve. 
Near the irregular singularity, the curve has the following form (we put the singularity at $z=\infty$):
\begin{align}
& \text{Type I}:~~(x^{N/d}+z^{k/d})^d=0, \nonumber\\
& \text{Type II}:~~x(x^{(N-1)/d}+z^{k/d})^d=0, \nonumber\\
& \text{Type III}:~~(x+z^{n-2})^N=0,
\end{align}
where in the Type I (or type II) case, $d$ is the maximal common divisor of $k$ and $N$ (or $N-1$), and we have neglected all the coefficients.
From these, one can easily see the rules stated in section~\ref{sec:2}.

\subsection{Local moduli from regular singularities}

Regular punctures are realized~\cite{Chacaltana:2012zy,Yonekura:2013mya} by 
coupling the 3d $\CN=4$ $T^\rho[SU(N)]$ theories~\cite{Gaiotto:2008ak} to the 5d SYM. These theories
are extended in the ${\mathbb R}^3$ direction, and they are localized at punctures on the Riemann surface $C$.

\subsubsection{$T^\rho[SU(N)]$ theories}
We first review some properties of $T^\rho[SU(N)]$ necessary for our purposes; see \cite{Gaiotto:2008ak,Chacaltana:2012zy} for more details.

\paragraph{Moduli space of $T[SU(N)]$ theory.}
First, let us recall some of the important properties of the $T[SU(N)]$ theory, i.e., $\rho=0$.
It has Coulomb branch and Higgs branch, and there are $SU(N)_\CC \times SU(N)_\CH$ global symmetries
where $SU(N)_\CC$ acts on the Coulomb branch and $SU(N)_\CH$ acts on the Higgs branch.
Since it has 3d $\CN=4$ supersymmetry, both of the branches are hyperkahler, and hence there are hyperkahler moment maps
$\vec{\nu}=(\nu_1,\nu_2,\nu_3)$ and  $\vec{\mu}=(\mu_1,\mu_2,\mu_3)$ of the flavor symmetries $SU(N)_\CC $ and $SU(N)_\CH$, respectively. 
They take values in the Lie algebra of $SU(N)_\CC \times SU(N)_\CH$.
It is convenient to define holomorphic moment maps as $\nu_h=\nu_1+i\nu_2$ and $\mu_h=\mu_1+i\nu_2$.
Then, they are nilpotent matrices, i.e., $\nu_h^N=\mu_h^N=0$, or equivalently their eigenvalues are zero, when there is no mass deformation.
(Mass deformation will be important later.) Vacuum moduli space is parametrized by $\mu_h$ and $\nu_h$.

Mixed branches of the $T[SU(N)]$ theory are given as follows. Let us take a partition of $N$, $Y'=[m'_a]=[m'_1,m'_2\cdots,m'_\ell]$.
Corresponding to each partition, we can define an embedding of $SU(2)$ into $SU(N)$,
\beq
\rho' : SU(2) \to SU(N),
\eeq
such that the fundamental ($N$-dimensional) representation of $SU(N)$ is decomposed into irreducible representations of $SU(2)$ as
\beq
{\bf N} \to {\bf m}'_1+{\bf m}'_2+\cdots+{\bf m}'_\ell,
\eeq
where ${\bf m}_a$ is the $m_a$ dimensional (spin $(m_a-1)/2$) representation of $SU(2)$.
Similarly, we can consider the dual Young diagram of $Y'$
which we call $Y'^D=[\tilde{m}'_a]$, and we can define another embedding ${\rho}'_D:SU(2) \to SU(N)$.

Define $e'=\rho'(\sigma^+)$ and ${e}'_D={\rho}'_D(\sigma^+)$, i.e., the image of the raising operator $\sigma^+$ of $SU(2)$.
Then, they are nilpotent matrices (because acting raising operators too many times gives zero). 
Mixed branches are classified by $Y'$, and the branch labeled by $Y'$ is given as
\beq
(\nu_h,\mu_h) \in  {O}_{{e}'_D} \times  {O}_{e'}
\eeq
where, for a given matrix $A$, $O_A$ is the orbit of $A$ by the action of the complexified group $SU(N)_{\mathbb C}=SL(N)$,
$ 
O_A=\{ g A g^{-1}: g \in SU(N)_{\mathbb C} \}.
$ 
This means that the mixed branch structure $\bigcup_\alpha \CC_\alpha \times \CH_\alpha$ of the $T[SU(N)]$ theory is given as
$\alpha=Y'$, $ \CC_\alpha={O}_{{e}'_D}$ and $\CH_\alpha={O}_{e'}$.

Let us compute the dimension of the orbit $O_{e'}$. It is computed as the number of linearly independent generators of $SU(N)$ which do not commute with
$e'=\rho'(\sigma^+)$. The adjoint representation of $SU(N)$ is decomposed under $SU(2)$ as
\beq
{\bf N^2-1} &\to \left[\bigoplus_{a=1}^{\ell} \bigoplus_{j=1}^{m'_a - 1} {\bf (2j+1)} \right] \bigoplus  \left[(\ell-1) \cdot {\bf 1} \right] \bigoplus 
 \left[ 2 \bigoplus_{a<b} \bigoplus_{k=1}^{m'_b} {\bf (m'_a+m'_b-2k+1)} \right] \nonumber \\
&\equiv \bigoplus_c {\bf (2j_c+1)}, \label{eq:adjdecompose}
\eeq
where $c$ in the last equation runs over irreducible representations which appear in the decomposition. 
Corresponding to this decomposition, the generators of $SU(N)$ are given as $T'_{c, m}$, $-j_c \leq m \leq j_c$ such that
$[\rho(\sigma^3/2),T'_{c,m}] =mT'_{c,m}$,~ $[\rho(\sigma^+),T'_{c,m}] \propto T'_{c,m+1}$ etc.
Generators corresponding to the highest spin state in each irreducible representation commutes with $\rho(\sigma^+)$, i.e., $[\rho(\sigma^+),T'_{c,j_c}]=0$.
So the complex dimension of $O_{e'}$ is given as
\beq
\dim_{\mathbb C} O_{e'} &=(N^2-1)-\left( \sum_{a=1}^{\ell}(m'_a-1)+(\ell-1)+  2\sum_{b=1}^\ell (b-1) m'_b \right) \nonumber \\
&=N^2+N-2\sum_{a} a m'_a=\left(N^2- \sum_{a} \tilde{m}'^2_a\right) \equiv 2d(Y'),
\eeq
where we used the relation between $Y'=[m'_a]$ and $Y'^D=[\tilde{m}'_a]$. Similarly, the complex dimension of $O_{{e}'_D}$ is given by 
$\dim_{\mathbb C} O_{{e}'_D}=2d(Y'^D)$.

\paragraph{Moduli space of $T^\rho[SU(N)]$ theory.}
An easy way to obtain the $T^\rho[SU(N)]$ is as follows. Take an embedding $\rho : SU(2) \to SU(N)$ corresponding to a 
partiton of $N$, $Y=[m_a]$.  
In the $T[SU(N)]$ theory, we give a vev $\vev{\mu_h}=\rho(\sigma^+) \equiv e$ to the moment map $\mu_h$.
Then, in the low energy limit,
the $T[SU(N)]$ theory flows to the $T^\rho[SU(N)]$ theory with some Goldstone multiplets associated to the 
spontaneous breaking of the flavor $SU(N)_\CH$ symmetry by the vev $\vev{\mu_h}=e$. By Goldstone multiplets, we mean
Goldstone bosons and their superpartners.

The Goldstone multiplets are given by the orbit of $e=\rho(\sigma^+)$.
Eliminating them, we get 
\beq
\mu_h&=\rho(\sigma^+)+\sum_{c,m} T_{c, m} \mu_{c,m} \nonumber \\
 &\to \rho(\sigma^+)+\sum_{c} T_{c, -j_c} \mu_{c,-j_c}
\label{eq:slodowy}
\eeq
where $T_{c,m}$ is defined according the decomposition as in \eqref{eq:adjdecompose} using $Y=[m_a]$.
In the above, we used the fact that $\mu_{c,m}$ for $m>-j_c$ are Goldstone multiplets and we have eliminated them.
Therefore, the Higgs branch of the $T^\rho[SU(N)]$ theory is parametrized by the fields $\mu_{c,-j_c}$.
A matrix of the form \eqref{eq:slodowy} is an element of the so called Slodowy slice $\CS_\rho$. 
The Higgs branch of the $T^\rho[SU(N)]$ theory is parametrized by this $\mu_h$ in \eqref{eq:slodowy}.

A matrix of the form \eqref{eq:slodowy} can take values in the orbit $O_{e'}$ for some $e'=\rho'(\sigma)$
if and only if $e$ can be reached as as a limit of elements of $O_{e'}$.
In other words, $e$ must be contained in the closure of $O_{e'}$, denoted as $\bar{O}_{e'}$.
So we must have $O_{e} \subset \bar{O}_{e'}$. 
It is known that $O_{e} \subset \bar{O}_{e'}$ if and only if the corresponding partitions $Y=[m_a]$ and $Y=[m'_a]$
satisfy $Y \leq Y'$, where the partial ordering between partitions of $N$ is defined in \eqref{eq:partitionorder}.

Now we can describe the mixed branch structure $\bigcup_\alpha \CC_\alpha \times \CH_\alpha$ of the $T^\rho[SU(N)]$ theory.
The branches are labeled by a partition $Y'=[m'_a]$ of $N$, which satisfies $Y' \geq Y$.
The Coulomb and Higgs components are given as
\beq
\CC_\alpha(T^\rho[SU(N)]) &=O_{{e}'_D} \\
\CH_\alpha(T^\rho[SU(N)]) &=O_{e'} \cap S_\rho,
\eeq
where $e'$ and ${e}'_D$ are the nilpotent elements corresponding to $Y'=[m'_a]$ and the dual $Y'^D=[\tilde{m}'_a]$ as before, 
and $S_\rho$ is the Slodowy slice \eqref{eq:slodowy}.
The quaternionic dimensions (or half the complex dimension) of them are given as 
\beq
\dim_{\mathbb H} \CC_\alpha(T^\rho[SU(N)]) &= \frac{1}{2} \dim_{\mathbb C} O_{{e}'_D}=d(Y'^D). \\
\dim_{\mathbb H} \CH_\alpha(T^\rho[SU(N)]) &= \frac{1}{2} \left( \dim_{\mathbb C} O_{e'}- \dim_{\mathbb C} O_{e} \right)=d(Y')-d(Y).
\eeq
The subtraction of $\dim_{\mathbb C} O_{e}$ in $\dim_{\mathbb H} \CH_\alpha$ comes from the fact that 
we have eliminated the Goldstone multiplets in \eqref{eq:slodowy}.

For example, the pure Coulomb branch $\CC \times \{0 \}$ is given by the orbit $O_{e_D}$ corresponding to the partition $Y^D=[\tilde{m}_a]$
dual to $Y=[m_a]$. This will reproduce the usual rule of regular singularities given in \eqref{eq:degofsing}.

Notice that the constraint $Y' \geq Y$ is equivalent to $Y'^D \leq Y^D$.\footnote{
If two partitions $[\tilde{m}_a]$ and $[\tilde{m}'_a]$ satisfy $[\tilde{m}'_a] \leq [\tilde{m}_a]$, then 
$[m'_a] \geq [m_a]$.  
This claim is proved as follows. Examining the Young diagram, one can see that
$\sum_{b=1}^a m'_b=N- \sum_{b} \max \{ \tilde{m}'_b-a,0 \}$. Then by taking $c$ such that $\tilde{m}'_c=a$ and $\tilde{m}'_{c+1}<a$,
we get $\sum_{b} \max \{ \tilde{m}'_b-a,0 \}=\sum_{b=1}^c ( \tilde{m}'_b-a) \leq \sum_{b=1}^c ( \tilde{m}_b-a) \leq \sum_{b} \max \{ \tilde{m}_b-a,0 \}$,
where we have used the definition of the partial ordering \eqref{eq:partitionorder}. 
Thus we get $\sum_{b=1}^a m'_a \geq \sum_{b=1}^a m_a$ for $a=1,2,\cdots$,
proving $[m'_a] \geq [m_a]$. }
So in all the branches, we must have the constraint
\beq
\nu_h \in \bar{O}_{{e}_D}. \label{eq:nuconstraint}
\eeq

\subsubsection{$T^\rho[SU(N)]$ coupled to 5d SYM}
Now let us couple the $T^\rho[SU(N)]$ theory to the 5d SYM at a point $p \in C$ which will be a puncture.
We gauge the Coulomb branch $SU(N)_\CC$ symmetry of $T^\rho[SU(N)]$ by the gauge group of the 5d SYM.
In particular, this coupling includes a superpotential coupling between the holomorphic moment map $\nu_h$ and the complex adjoint field
$\varphi_h=\varphi_1+i\varphi_2$,
\beq
W \supset \tr(\varphi_h(p) \nu_h),\label{eq:supercoupling}
\eeq
where $\varphi_h(p)$ is the value of $\varphi_h$ evaluated at the point $p$. 

There are two important effects of this coupling.
First, by giving the vev \eqref{eq:constdiagvev}, this coupling gives a mass term of $\nu_h$.
Then the Higgs branch moment map $\mu_h$ is no longer a nilpotent matrix, 
but it is such that its characteristic polynomial is given by~\cite{Yonekura:2013mya}
\beq
\det (x I_N-\mu_h)=\det (x I_N-\varphi_h), \label{eq:muconstraint}
\eeq
where $x$ is an arbitrary variable. (When $\varphi_h=0$, this equation means that all the eigenvalues of $\mu_h$ are zero, and hence
$\mu_h$ is nilpotent.) 

The other effect of \eqref{eq:supercoupling} is that this term gives a source of $\Phi$.
In the bulk superpotential of the 5d SYM, there is also a coupling of the form $W \supset \int d^2z \tr(\varphi_hD_{\bar{z}} \Phi)$ and the equation
of motion of $\varphi_h$ gives $D_{\bar{z}}\Phi \sim \delta^{2}(z) \nu_h$, where we have assumed that the point $p$ is located at $z=0$.
Then we get a pole of $\Phi$ as \footnote{
Strictly speaking, \eqref{eq:hitchinpole} is not an actual solution of Hitchin's equations if $[\nu_h, \nu_h^\dagger] \neq 0$. 
We must use the usual argument that
imposing \eqref{eq:bps1} and dividing by the gauge group is equivalent to dividing by complexified gauge group. Then,
using complexified gauge transformation to set $A_{\bar{z}}=0$, we get \eqref{eq:hitchinpole}. }
\beq
\Phi \sim \frac{\nu_h}{z}+\CO(1)~~~(z \to 0).\label{eq:hitchinpole}
\eeq
Suppose that the $\nu_h$ is in the orbit $\nu \in O_{{e}_D}$ corresponding to the partition $[\tilde{m}_a]$.
Then, combining the pole term $\nu_h/z$ and also the order one term $\CO(1)$ in the above equation, one can check 
that the singularity of the spectral curve is precisely given by \eqref{eq:degofsing}.
The constraint \eqref{eq:nuconstraint} ensures that the singularities cannot be stronger than \eqref{eq:degofsing}.

Now we can study the local contribution $\CC_\alpha(p) \times \CH_\alpha(p)$ from the puncture $p$
to the mixed branch $\CC_\alpha \times \CH_\alpha$ of the 4d field theory.

\paragraph{Coulomb factor.}
The commutation relation $[\Phi, \vec{\varphi}]=0$ requires that the residue $\nu_h$ of the pole of $\Phi$ at the puncture as given in \eqref{eq:hitchinpole} 
must be of the form
\beq
\nu_h=\left( \begin{array}{cccc}
\nu_h^{(1)} & 0 & \ldots  &0 \\
0 & \nu_h^{(2)} & & \vdots \\
\vdots& & \ddots &  0\\
0&\ldots& 0 & \nu_h^{(k)}
\end{array}
\right) . \label{eq:blocknu}
\eeq
Furthermore, each of the block $\nu_h^{(s)}$ must be nilpotent. Thus there is a partition of $n_s$, $Y^D_s=[\tilde{m}_{s,a}]$,
the corresponding embedding ${\rho}_D^{(s)}:SU(2) \to SU(n_s)$ and the nilpotent element ${e}(s)_D={\rho}^{(s)}_D(\sigma^+)$
such that 
\beq
\nu_h^{(s)} \in O_{{e}(s)_D}~~~(s=1,2,\cdots,k). \label{eq:blocknilp}
\eeq
Then, the singularity of the spectral curve is specified by $Y^D_s$.

Let us combine the partitions $Y^D_s$ to define $Y'^D=[\tilde{m}'_a]$ whose columns are the columns of $Y^D_s$.
Because of the constraint \eqref{eq:nuconstraint}, this partition must satisfy $Y'^D \leq Y^D$.

The Coulomb factor $\CC_\alpha(p)$ is parametrized by $\nu_h^{(s)}$. However, the following point must be taken into account.
As explained in section~\ref{sec:bulk}, we compactified the 4d theory on $S^1$. Then, in the low energy 3d theory,
a $U(1)$ massless gauge multiplet is dualized to two real scalars (or equivalently one complex scalar).
Therefore, the dimension of Coulomb moduli space $\CC_\alpha^{(3d)}$ is doubled from the 4d Coulomb moduli space $\CC_\alpha^{(4d)}$.
Having this distinction between the 3d and 4d theory in mind, we conclude that
\beq
\CC_\alpha^{(3d)}(p)=\bigoplus_{s=1}^k O_{{e}(s)_D}
\eeq
and 
\beq
\dim_{\mathbb C} \CC_\alpha^{(4d)}(p)=\dim_{\mathbb H} \CC_\alpha^{(3d)}(p)=\sum_{s=1}^k d(Y_s^D).
\eeq
This is what we have written in \eqref{eq:localCdim}.

\paragraph{Higgs factor.}
The $\mu_h$ must satisfy \eqref{eq:muconstraint}. Furthemore, the $\nu_h$ satisfies \eqref{eq:blocknu} and \eqref{eq:blocknilp}.
From these facts, it is natural to expect the following (see section~2.3 and 2.4 of \cite{Chacaltana:2012zy}).
For each $s$, let $Y_s=[m_{s,a}]$ be the partition of $n_s$ dual to $Y^D_s$, $\rho^{(s)}:SU(2) \to SU(n_s)$ be the corresponding
embedding, and $e(s)=\rho^{(s)}(\sigma^+)$. Let $e''=\bigoplus_s e(s)$ be the block diagonal $N \times N$ matrix.
Then we define the matrix
\beq
\varphi_h+e''= \left( \begin{array}{cccc}
a_h^{(1)} I_{n_1} + e(1)& 0 & \ldots  &0 \\
0 & {a}_h^{(2)} I_{n_2} +e(2)& & \vdots \\
\vdots& & \ddots &  0\\
0&\ldots& 0 & {a}_h^{(k)} I_{n_k} +e(k)
\end{array}
\right), \label{eq:deformedmat}
\eeq
where $a_h^{(s)}=a_1^{(s)}+ia_2^{(s)}$. Suppose that $a_h^{(s)}$ are generic so that $a_h^{(s)} \neq a_h^{(t)}$ for $s \neq t$.
In this case, we propose that $\mu_h$ is in the orbit of $\varphi_h+e''$, 
\beq
\mu_h \in \CH_\alpha(p) \equiv O_{\varphi_h+e''} \cap S_\rho.
\eeq
The complex dimension of the orbit $O_{\varphi_h+e''}$ is given by
\beq
\dim_{\mathbb C} O_{\varphi_h+e''}&=\dim_{\mathbb C} O_{\varphi_h}+\sum_{s=1}^k \dim_{\mathbb C} O_{e(s)} \nonumber \\
&=\left(N^2-\sum_{s=1}^k n_s^2 \right)+\sum_{s=1}^k \left( n_s^2- \sum_a \tilde{m}_{s,a}^2\right) \nonumber \\
&=2d(Y')
\eeq
where $Y'=[m'_a]$ is the partition of $N$ dual to $Y'^D=[\tilde{m}'_a]$.

We can perform a consistency check by taking the limit $\varphi_h \to 0$.
In this case, the $T^\rho[SU(N)]$ theory at the puncture has no (holomorphic) mass term, and hence we have
\beq
\mu_h \in O_{e'} \cap S_\rho,
\eeq
where $e'=\rho'(\sigma^+)$ is a nilpotent element corresponding to the partition $Y'=[m'_a]$.
Therefore, we should get
\beq
\lim_{\varphi_h \to 0} O_{\varphi_h+e''}=O_{e'}.
\eeq
Actually, the dimension of $O_{\varphi_h+e''}$ computed above is the same as the dimension of $O_{e'}$.
This is required because, even if $\varphi_h =\varphi_1+i\varphi_2 \to 0$, this limit should be smooth as long as $\varphi_3 $ is generic.\footnote{
The $\varphi_3$ does not appear in the complex structure of $T^\rho[SU(N)]$, but it appears in the metric.
If we use $SU(2)_R$ symmetry, the case $\varphi_h=0$ and $\varphi_3$=generic can be transformed into $\varphi_h$=generic.}

We conclude that the quaterionic dimension of $\CH_\alpha(p)$ is given as
\beq
\dim_{\mathbb H} \CH_\alpha(p)=d(Y')-d(Y).
\eeq
This is the formula written in \eqref{eq:localHdim}.

\section{$\CN=2$ SUSY QCD} \label{sec:4}

In this section, we reproduce the results of the moduli space of vacua in SQCD
obtained by Seiberg and Witten for $SU(2)$~\cite{Seiberg:1994aj} and by Argyres, Plesser and Seiberg for $SU(N)$~\cite{Argyres:1996eh}.

\subsection{$SU(2)$}
We study the $SU(2)$ SQCD with $N_f$ flavors.
Although these models are simple, they illustrate our method without technical complication.

We consider massless cases. In addition to the regular singularities discussed in the previous sections, we also need
some irregular singularities. The curve is $x^2+\phi_2(z)=0$, and the singularities we use are given as
\beq
A:&~ \phi_2 = \CO(z^{-1}) , \label{eq:SU2A} \\
B:&~ \phi_2=\frac{\Lambda^2}{z^3}+\CO(z^{-2}), \label{eq:SU2B} \\
C:&~ \phi_2=\frac{\Lambda^2}{z^4} +\CO(z^{-2}), \label{eq:SU2C}
\eeq
where $\Lambda$ is a parameter which roughly corresponds to the dynamical scales of field theory.
The first singularity A is the regular singularity corresponding to the partition $Y=[m_a]=[1,1]$ or equivalently its dual partition is 
$Y^D=[\tilde{m}_a]=[2]$. 
The second and third singularities B and C are irregular singularities
which are obtained~\cite{Gaiotto:2009hg} from the M-theory uplift of type IIA brane configurations~\cite{Witten:1997sc}.
More details on those types of irregular singularities are discussed in appendix~\ref{sec:appA}, 
but here we simply claim that \eqref{eq:SU2B} and \eqref{eq:SU2C}
have no local contribution to the Higgs branch.

Let us summarize the important points.
\begin{itemize}
\item If the curve is factorized as  $x^2+\phi_2=(x+f(z))(x-f(z))$, there is a contribution to the quaternionic dimension of the Higgs branch as $+1$
coming from the term $k-1$ in \eqref{eq:higgsdim}, i.e., the curve is factorized according to the partition $X=[n_s]=[1,1]$.
\item If there is a singularity of type B, then it is impossible to factorize the curve as $x^2+\phi_2=(x+f(z))(x-f(z))$
since we cannot have
$\Lambda^2/z^3+\cdots=(f(z))^2$ for some single valued function $f(z)$. On the other hand, in the case of the singularity of type C, there is
no such local obstacle.

\item For a singularity of type A, we can have local contribution to the Higgs branch if and only if the $\CO(z^{-1})$ term vanishes.
In this case we have $Y'^D=[\tilde{m}'_a]=[1,1]$, and
the local contribution to the quaternionic dimension, calculated by the formula \eqref{eq:localHdim}, is $\frac{1}{2}(2^2-1^2-1^2)=1$.

\end{itemize}

\paragraph{Zero flavor $N_f=0$.}
The theory with no flavor is realized by putting two irregular singularities of type B, \eqref{eq:SU2B}, on a Riemann sphere parametrized by
$z \in {\mathbb C} \cup \{ \infty \}$.
Putting the singularities at $z=0$ and $z=\infty$, the curve is
\beq
0=x^2-\left(\frac{\Lambda^2}{z^3}+\frac{u}{z^2}+\frac{\Lambda^2}{z} \right),
\eeq
where $u$ is the Coulomb modulus. (Note that near $z \to \infty$, we must use the coordinates $z'=1/z$ and $x'=-z^2 x$.)
We cannot factorize the curve, and there are no local contributions to the Higgs branch from the punctures. Thus there is no Higgs branch,
reproducing the result of field theory.

\paragraph{One flavor $N_f=1$.}
The theory with one massless flavor is realized by putting one irregular singularity of type B, \eqref{eq:SU2B}, and one irregular singularity of 
type C, \eqref{eq:SU2C}, on a Riemann sphere. Putting singularities at $z=0$ and $z=\infty$, the curve is
\beq
0=x^2-\left(\frac{\Lambda^2}{z^4}+\frac{u}{z^2}+\frac{\Lambda^2}{z} \right).
\eeq
In this case, there is no Higgs branch as in the case of $N_f=0$.

\paragraph{Two flavors $N_f=2$.}
The theory with two massless flavors can be realized in two different ways. The first way is to put
two type C singularities. The second way is to put one type B irregular singularity and two type A singularities, \eqref{eq:SU2A}.

First, let us study the first realization.
We use two type C singularities. Putting singularities at $z=0$ and $z=\infty$, the curve is
\beq
0=x^2-\left(\frac{\Lambda^2}{z^4}+\frac{u}{z^2}+\Lambda^2 \right).
\eeq
The curve can be factorized if and only if $u=\pm 2\Lambda^2$. In this case, we have
\beq
0=\left(x+ \left( \frac{\Lambda}{z^2} \pm \Lambda \right)\right)\left(x- \left(\frac{\Lambda}{z^2} \pm \Lambda \right)\right).
\eeq
Therefore, the Higgs branch is emanating from the Coulomb branch points $u=\pm 2\Lambda^2$ and its quaternionic dimension is $1$. 
This is exactly as was found by Seiberg and Witten~\cite{Seiberg:1994aj}.

Next, let us study the second realization. We take regular punctures of type A at $z= 0,1$ and a puncture of type B at $z=\infty$. The curve is
\beq
0=x^2-\left( \frac{\Lambda^2}{z} +\frac{u'}{z(z-1)} \right).
\eeq
We cannot factorize the curve in this case because of the existence of the singularity of type B. 
However, there are contributions to the Higgs branch dimension from regular punctures.
Notice that
\beq
\left( \frac{\Lambda^2}{z} +\frac{u'}{z(z-1)} \right) \to \left\{
\begin{array}{ll}
\displaystyle{\frac{\Lambda^2-u'}{z} }&~~~ z \to 0 \\[0.3cm]
\displaystyle{\frac{u'}{z-1} }&~~~ z \to 1
\end{array}
\right.
\eeq
Then, we can have local contribution to the Higgs branch from the puncture at $z=0$ when $u'=\Lambda^2$, and from the puncture at $z=1$ when $u'=0$.
At each of these points, we have the Higgs branch with the quaternionic dimension $1$. Therefore, we find the same structure
as the case of the above first realization if we identify the moduli parameters as $u=2u'-\Lambda^2$.

It is interesting that the way the Higgs branch appears is quite different in the two realizations. 
In the first realization, the Higgs branch comes from the ``bulk'' contribution, while in the second realization the Higgs branch comes from local 
contributions from punctures. However, the final results are the same in both cases.

\paragraph{Three flavors $N_f=3$.}
The theory with three massless flavors is realized by putting two regular singularities of type A at $z=0,1$ 
and one irregular singularity of type C at $z=\infty$. The curve is
\beq
0=x^2-\left( \Lambda^2 +\frac{u}{z(z-1)} \right).
\eeq
Note that the local contributions to the Higgs branch from the punctures at $z=0,1$ are both turned on when $u=0$.
Furthermore, only in this case the curve is factorized,
\beq
0=\left(x+\Lambda^2\right)\left(x-\Lambda^2\right).
\eeq
Therefore, the Higgs branch is emanating from the single point $u=0$ and the quaternionic dimension is 3.
This is exactly as in \cite{Seiberg:1994aj}.

\paragraph{Four flavors $N_f=4$.}
The theory with four massless flavors is realized by putting four regular singularities of type A at $z=0,q,1,\infty$. 
The curve is
\beq
0=x^2- \left(\frac{u}{z(z-q)(z-1)} \right).
\eeq
All the local contributions to the Higgs branch and the factorization of the curve is possible if and only if $u=0$.
Therefore, the Higgs branch is emanating from $u=0$ and has the quaternionic dimension 5.
This point is the superconformal point of the theory.

\paragraph{Mass deformation and AD point.}
Let us briefly describe mass deformation.
Generic deformation would typically eliminate the Higgs branch. Consider the first realization of $SU(2)$
theory with two flavors. The curve turning on generic mass deformation is
\begin{equation}
x^2={\Lambda^2\over z^4}+{m_1\Lambda\over z^3}+{u\over z^2}+{m_2 \Lambda \over z}+\Lambda^2. 
\end{equation}
The above curve can be factorized if and only if
\begin{align}
& m_1=m_2=m,~~~~{u\over \Lambda^2}={m^2\over 4\Lambda^2}+2, \nonumber \\
& m_1=-m_2=m,~~{u\over \Lambda^2}={m^2\over 4\Lambda^2}-2. \label{eq:fconditiononm}
\end{align}
So for non-zero $m=m_1=\pm m_2$, there is only one Higgs branch of dimension 1, which is consistent with the fact that
the flavor symmetry is broken to $U(2)$ by the mass term. 
The above curve can be further factorized for the following special value of $m$:
\begin{align}
& x^2={\Lambda^2\over z^4}(z\pm1)^4,~~m_1=m_2=m=\pm 4\Lambda,  \nonumber\\
& x^2={\Lambda^2\over z^4}(z\pm i)^4,~~m_1=-m_2=m=\pm 4i\Lambda.
\end{align}
There should be new massless particles here, but our method tells us that the Higgs branch dimension does not change at this point. 
This suggests that the new massless particles might be nonlocal relative to the massless particles which are already present in \eqref{eq:fconditiononm}. 
This point is actually the Argyres-Douglas point as discussed in \cite{Argyres:1996hc}.

\subsection{$SU(N)$}
In this section we study the $SU(N)$ SQCD. First, let us summarize the result of Argyres, Plesser and Seiberg~\cite{Argyres:1996eh}
obtained by field theory methods.
\begin{itemize}
\item There is a baryonic Higgs branch for $N_f\geq N$, and the quaternionic dimension is $N(N_f-N)+1$. There is no Coulomb direction in this branch.
\item There are  non-baryonic mixed branches with a label $0 \leq r\leq \min([{N_f\over 2}], N-2)$. The complex dimension of the Coulomb
factor $\CC_r$ is $N-r-1$, and the quaternionic dimension of the Higgs factor $\CH_r$ is $r(N_f-r)$. 
The pure Coulomb branch is the case $r=0$.
\end{itemize}

Here we only treat the conformal theory $N_f=2N$ which only requires regular singularities.
Other cases are discussed in appendix~\ref{sec:appA}.

\subsubsection{$N_f=2N$}
This theory is described by a Riemann sphere with two simple punctures ($Y=[m_a]=[N-1,1]$ or $Y^D=[\tilde{m}_a]=[2,1^{N-2}]$) 
and two full punctures  ($[Y]=[m_a]=[1^N]$ or $Y^D=[\tilde{m}_a]=[N]$). 
We put two full punctures at $z=0,\infty$, and two simple punctures at $z=1,q$. 
The Seiberg-Witten curve is 
\begin{align}
0=x^N+\sum_{i=2}^{N}{u_i\over z^{i-1} (z-1)(z-q)}x^{N-i}.
\end{align}

\paragraph{Non-baryonic branch.}
Suppose that some of the $u_i$ is nonzero.
Then we can define $r$ ($0 \leq r \leq N-2$) such that $u_{N-r} \neq 0$ and $u_{N-r+1}=\cdots=u_N=0$. Then the curve is factorized as
\beq
0=x^r\left(x^{N-r}+\sum_{i=2}^{N-r}{u_i\over z^{i-1} (z-1)(z-q)}x^{N-i} \right).
\eeq
This means that we have factorization corresponding to the partition $X=[n_s]=[N-r,1^r]$. 

\begin{center}
\begin{figure}[htbp]
\small
\centering
\includegraphics[width=12cm]{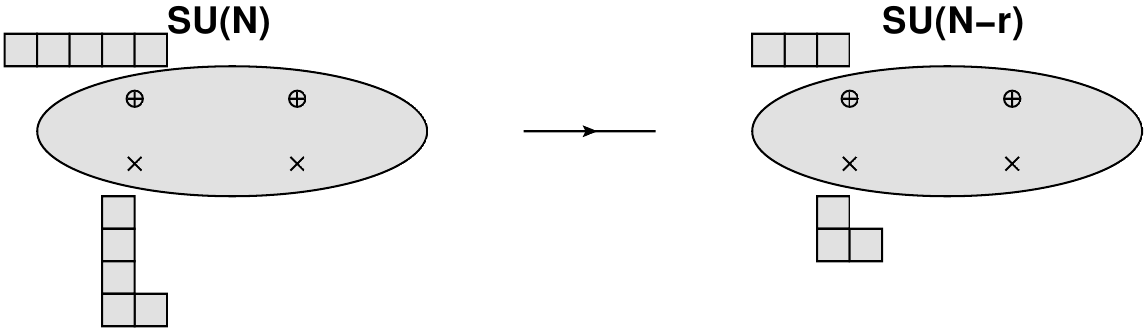}
\caption{The factorization patter of $r$th mix branch of the $SU(N)$ theory with $2N$ flavors.}
\label{split}
\end{figure}
\end{center}

In the second step, we should choose 
the local puncture type on the $SU(N-r)$ factor. However, it is easy to see that to have 
nonzero $u_{N-r}$, the punctures  have to be chosen as in the figure \ref{split}: the full (simple) puncture of $SU(N)$ becomes the full (simple) 
puncture of $SU(N-r)$.
We have the following data on the split of punctures:
\begin{align}
&\text{Full puncture}:~~~~~Y_f^D=[N],~~~~~~~~~~~Y_f^{'D}=[N-r,1^{r}] \nonumber\\
&\text{Simple puncture}:~Y_s^D=[2,1^{N-2}],~~~~Y_s^{'D}=[2,1^{N-2}]
\end{align}

Using the formula \eqref{eq:localHdim}, local contribution to the Higgs component from a full or simple puncture is given by
\beq
&\dim_{\mathbb H} \CH_r({\rm full})=d(Y_f^{'})-d(Y_f)=\frac{1}{2}\left(N^2-(N-r)^2-r \cdot 1^2 \right) \\
&\dim_{\mathbb H} \CH_r({\rm simple})=d(Y_s^{'})-d(Y_s)=0.
\eeq
Then, the formula \eqref{eq:higgsdim} gives the total Higgs branch dimension as
\beq
\dim_{\mathbb H} \CH_r&=r+2\dim_{\mathbb H} \CH_r({\rm full})+2\dim_{\mathbb H} \CH_r({\rm simple}) \nonumber \\
&=r(2N-r),
\eeq
where we have used $k=r+1$.
This is exactly what was found in \cite{Argyres:1996eh} with $N_f=2N$.

\paragraph{Baryonic branch.}
In the above we have assumed $u_{N-r} \neq 0$ for some $r$. The remaining possibility is to have all the $u_i$ to be zero.
The curve is just $0=x^N$, which corresponds to the partition $X=[n_s]=[1^N]$.
At all the punctures we automatically have $Y'^D=[\tilde{m}'_a]=[1^N]$.

Local contribution from a full or simple puncture is given by
\beq
&\dim_{\mathbb H} \CH({\rm full})=\frac{1}{2}\left(N^2-N \cdot 1^2 \right) \\
&\dim_{\mathbb H} \CH({\rm simple})=\frac{1}{2}\left(N^2-N \cdot 1^2 \right)-\frac{1}{2} \left(N^2-2^2-(N-2)\cdot 1^2 \right).
\eeq
Then, the formula \eqref{eq:higgsdim} gives
\beq
\dim_{\mathbb H} \CH&=(N-1)+2\dim_{\mathbb H} \CH_r({\rm full})+2\dim_{\mathbb H} \CH_r({\rm simple}) \nonumber \\
&=N^2+1.
\eeq
This is exactly what was found in \cite{Argyres:1996eh} with $N_f=2N$.

\section{Theories defined using regular singularities}\label{sec:5}

Our method can be also applied to theories whose Lagrangian description is not yet know.
One of such theories is the so called $T_N$ theory.
Pure Higgs branch $\CH$ (i.e., branch where there are no Coulomb moduli fields) are studied in \cite{Gaiotto:2008nz,Maruyoshi:2013hja}.
But it is not so easy to study the mixed branch structures directly by field theory.
Here we use our method to study these structures. The case of the partition $X=[n_s]=[1^N]$ was studied in \cite{Yonekura:2013mya}.

\subsection{$T_N$ theory}

The $T_N$ theory is realized by using a Riemann sphere with three full punctures.
Each full puncture has $SU(N)_\CH$ Higgs branch symmetry, so the $T_N$ theory has $SU(N)_A \times SU(N)_B \times SU(N)_C$ flavor symmetry.

In this case, what are discussed in sections~\ref{sec:2} and \ref{sec:3} suggest the following interpretations.
First, the factorization of the curve \eqref{eq:facto} is realized by giving vevs to Higgs branch operators of the $T_N$ theory so that the low energy theory
is given by smaller $T_N$ theories~\cite{Maruyoshi:2013hja}
\beq
T_N \to T_{n_1} +T_{n_2}+\cdots+T_{n_k}+({\rm free~fields}).
\eeq
Here the free fields come from some $k-1$ moduli fields responsible for the above breaking which roughly represent the relative ``positions'' of
$T_{n_s}$ on moduli space,\footnote{In terms of the M5 brane realization of the $T_N$ theory, this means that the $N$ M5 branes are separated as
$N \to [n_1,n_2,\cdots,n_s]$.}
and also the Goldstone multiplets associated to the spontaneous breaking $[SU(N)]^3 \to [{ S}(U(n_1) \times \cdots \times U(n_k) )]^3$.
There are $3 \times \frac{1}{2} (N^2-\sum_{s=1}^k n_s^2)$ Goldstone multiplets (in quaternionic dimension).

Next, we consider the partition $n_s \to Y^D_{s}=[\tilde{m}_{s,a}]$ at three punctures, which we call $A,B$ and $C$ respectively.
Thus, for each $s$, we have $Y^D_{s,A},Y^D_{s,B}$ and $Y^D_{s,C}$, and their
dual partitions $Y_{s,A},~Y_{s,B}$ and $Y_{s,C}$.
Then, we can Higgs the punctures of the $T_{n_s}$ theory by giving nilpotent vevs to the moment maps to get 
more general three-punctured theories specified by these partitions~\cite{Benini:2009gi,Chacaltana:2012zy,Tachikawa:2013kta}
\beq
T_{n_s} \to T_{n_s}\left(Y_{s,A},Y_{s,B},Y_{s,C}\right)+({\rm Goldstone~multiplets}),
\eeq
where the Goldstone multiplets are the ones associated to the spontaneous breaking of $[SU(n_s)]^3$ by nilpotent vevs.
As discussed in section~\ref{sec:3}, each puncture contributes $d(Y_{s})=\frac{1}{2}(n_s^2-\sum_a \tilde{m}_{s,a}^2)$ 
to the quaternionic dimension of Goldstone multiplets.
Combining all the low energy free fields in the above process gives the dimension of the Higgs factor $\CH_\alpha$ given in \eqref{eq:higgsdim}.
Then we go to the Coulomb branch of the remaining theory
\beq
\sum_{s=1}^{k} T_{n_s}\left(Y_{s,A},Y_{s,B},Y_{s,C}\right).
\eeq
This process gives the mixed branches of the $T_N$ theory.

For this process to define a mixed branch of the $T_N$ theory, $T_{n_s}\left(Y_{s,A},Y_{s,B},Y_{s,C}\right)$
must be such that its generic curve cannot be factorized. 
If some of $n_s$ is 1, we interpret the $T_{n=1}$ theory to be empty.

The above picture of the vacua structure suggests that we can find all the lower three punctured sphere theory on the Coulomb branch moduli space 
of the $T_N$ theory.

\subsection{Maldacena-Nunez theory} 
$\CN=2$ Maldacena-Nunez theories~\cite{Maldacena:2000mw,Gaiotto:2009gz} are 
defined by a genus $g$ Riemann surface with no punctures at all.
Therefore, we only need to consider a partition $X=[n_s]$ of the curve $0=x^N+\sum_i \phi_i(z) x^{N-i}$.

If $g=0$, the only possible curve is just $0=x^N$, since there are no nontrivial holomorphic sections of $\phi_i$.
Hence the only branch allowed by our rule is the case $X=[1^N]$. This is the Higgs branch with quaternionic dimension $N-1$.
There is no Coulomb branch.

If $g=1$, then the most general curve is given by $0=\prod_{i=1}^N(x-c_i)$ where $c_i$ are constants with $\sum_{i=1}^N c_i=0$.
Therefore, also in this case, the only possible partition allowed is $X=[1^N]$. The quaterionic dimension of the Higgs branch is $N-1$.
In fact, the Coulomb branch and Higgs branch are combined into a manifold with complex dimension $3(N-1)$, which agrees with 
the field theory result as the theory is just $\mathcal{N}=4$ SYM.

If $g>1$, we can have the following factorization:
\beq
0=\prod_{s=1}^k \left(x^{n_s}+\sum_{i=1}^{n_s} \phi_{s,i}(z)x^{n_s-i} \right),
\eeq
This means all the partitions $[n_a]$ are possible, and one can easily find the dimensions of the Higgs and Coulomb factor using 
our general formula. 

Let's remark an interesting point for the branch labeled by  the partition $X=[1^N]$. In this branch, we have the maximal Higgs branch deformations.
Unlike the sphere case, we still have Coulomb branch deformations which are described by the curve
\begin{equation}
(x-h_1)(x-h_2)\ldots(x-h_N)=0,
\end{equation}
and $h_i$ are sections of the canonical bundle. This means that there is no pure Higgs branch for this class of theories. 
See \cite{Yonekura:2013mya} for more detailed discussions on this case.

\section{Argyres-Douglas theory}\label{sec:6}

Here we use our method to determine the Higgs branch of Argyres-Douglas theory~\cite{Argyres:1995jj,Argyres:1995xn,Xie:2012hs}.
We will recover and extend the results obtained in \cite{Argyres:2012fu}.
The theories we consider are discussed in \cite{Xie:2012hs,Bonelli:2011aa,Cecotti:2010fi}.

\subsection{$(A_1, A_{N})$ theory}
This theory is described by a sphere with one irregular singularity which we put at $z=\infty$. 
\paragraph{N=2n-1.}
The Seiberg-Witten curve is 
\begin{align}
& x^2=z^{2n}+c_{2} z^{2n-2}+\cdots+c_{n} z^{n}+c_{n+1} z^{n-1}+u_{n+2} z^{n-2}+\ldots+u_{2n}
\end{align}
where $c_i~(i=2,\cdots,n)$ are coupling parameters of the theory, $c_{n+1} $ is a mass parameter, 
and $u_i~(i=n+2,\cdots,2n)$ are Coulomb branch moduli. 

To find the Higgs branch, we need to consider the factorization of the Seiberg-Witten curve. In fact, there is a factorization happening at a single point on the moduli space:
\beq
&x^2-(z^{2n}+c_{2} z^{2n-2}+\cdots+c_{n+1} z^{n-1}+u_{n+2} z^{n-2}+\ldots+u_{2n}) \nonumber \\
=& \left(x+z^n+\sum_{i=2}^na_i z^{n-i}\right)\left(x-z^n-\sum_{i=2}^na_i z^{n-i}\right). \label{eq:ANfacto}
\eeq
There are $2n-1$ equations for $2n-1$ parameters $c_{n+1}, u_i~(i=n+2,\cdots,2n)$ and $a_i~(i=2,\cdots,n)$ when $c_i~(i=2,\cdots,n)$
are fixed. By tuning $c_{n+1}$ appropriately,
factorization occurs for certain values of $u_i$ determined by $c_i$ uniquely. For example, for the SCFT case (${c_i=0}$), 
we find a one dimensional Higgs branch coming out from $u_i=0$.

\paragraph{N=2n.}

The curve for this theory is 
\begin{align}
x^2=z^{2n+1}+c_{2} z^{2n-1}+\cdots+c_{n+1} z^{n}+u_{n+2}z^{n-1}+\cdots+u_{2n+1} 
\end{align}
It is impossible to factorize the curve because of the existence of the leading term $z^{2n+1}$.
Therefore there is no Higgs branch.

\subsection{$(A_1, D_{N})$ theory}
This theory is described by one irregular singularity at $z=\infty$ and one regular singularity at $z=0$.
\paragraph{N=2n+2.}
The curve is
\beq
x^2=z^{2n}+c_1z^{2n-1}+\cdots+c_{n+1} z^{n-1}+u_{n+2}z^{n-2}+\cdots+{u_{2n+1}\over z}+{m^2\over z^{2}},
\eeq
where $c_i~(i=1,\cdots,n)$ are some parameters, $c_{n+1}$ and $m$ are mass parameters and $u_i~(i=n+2,\cdots,2n+1)$ are Coulomb moduli.

We only consider the case in which $m=0$ and also $c_{n+1}$ is tuned appropriately so that the curve can be factorized.
Then, if $u_{2n+1}=0$, we get a local contribution to the Higgs factor $\CH_\alpha$ from the puncture at $z=0$.
Therefore, we have a mixed branch at $u_{2n+1}=0$. The Coulomb factor $\CC_\alpha$ is the same as the $(A_1,A_{2n-1})$ theory discussed above,
and the Higgs factor $\CH_\alpha$ has quaternionic dimension 1.

At a single point on the Coulomb moduli space, we can factorized the curve as
\beq
&x^2-z^{2n}+c_1z^{2n-1}+\cdots+c_{n+1} z^{n-1}+u_{n+2}z^{n-2}+\cdots+{u_{2n+1}\over z} \nonumber \\
=&\left(x+z^n+\sum_{i=2}^na_i z^{n-i}\right)\left(x-z^n-\sum_{i=2}^na_i z^{n-i}\right).
\eeq
At this point, we must also set $u_{2n+1}=0$. There is a  
Higgs branch coming out of this point. The quaternionic dimension is $2$ by counting  
the contributions from the bulk and the puncture. 
In particular, the Higgs branch at the SCFT point ($a_i=0$) has dimension $2$, which agrees with the result in \cite{Argyres:2012fu}.

\paragraph{N=2n+3.} The curve is
\beq
x^2=z^{2n+1}+c_1z^{2n}+\cdots+c_{n+1} z^{n}+u_{n+2} z^{n-1}+\cdots+{u_{2n+2}\over z}+{m^2\over z^{2}},
\eeq
Again we consider the case $m=0$. Then, there is a mixed branch defined by $u_{2n+2}=0$.
The contribution to the Higgs factor $\CH_\alpha$ comes from the puncture at $z=0$,
and the Coulomb factor $\CC_\alpha$ is the same as the $(A_1,A_{2n})$ theory.

\subsection{$(A_1, E_{N})$ theory}
The Seiberg-Witten curve at the $E_N$ type SCFT  point is
\begin{align}
&E_6:~~~~~x^3+z^4=0, \nonumber \\
&E_7:~~~~~x^3+x z^3=0,\rightarrow x(x^2+z^3)=0, \nonumber\\
&E_8:~~~~~x^3+z^5=0.
\end{align}
It is easy to see that only the curve for $E_7$ theory is factorizable, and there is only one type of factorization with $X=[2,1]$, so 
the dimension of Higgs branch is one.

\section{Conclusion}\label{sec:7}
We have developed a general method to find the vacua structure of class ${\cal S}$ theory using generalized Hitchin equation.
We derived our results using the geometric method from six dimensional $(2,0)$ theory. 
For those 4d theories with lagrangian description, it  is interesting to check our result using field theory calculations. 
The holomorphic factorization of Seiberg-Witten curve and reduction of singularity at punctures
play a crucial role in our method, and a field theory interpretation of this 
fact would be welcome.

Let's summarize the possible moduli space structure of  ${\mathcal N}=2$ theories:
\begin{itemize}
\item Typically, there is a Coulomb branch, a Higgs branch and mixed branches.
Such examples include the theory defined by a sphere with regular punctures.
\item Sometimes there is only a Coulomb branch. The typical example is the pure $SU(2)$ SYM and
 $(A_{1}, A_{N-1})$ AD theory with even $N$.
\item There is only a Higgs branch for the free field theory such as bifundamentals. 
\item There is no pure Higgs branch, namely only a Coulomb and mixed branches exist. This situation happens e.g.
for the theory defined on a higher genus Riemann surface. 
\end{itemize}

We focused on the theories obtained from $A_{N-1}$ type $\CN=(2,0)$ theory in this paper, 
but our method is also applied to other theories using $D$ \cite{Tachikawa:2009rb,Chacaltana:2011ze} and $E$ \cite{Chacaltana:2014jba} type
 $(2,0)$ theory. It would be 
interesting to extend our study to those other types of theories.

The singularity structure on the Coulomb branch of $\mathcal{N}=2$ theory is crucial for the discovery of Seiberg-Witten solution 
\cite{Seiberg:1994rs,Seiberg:1994aj}. New massless particles appear at those singularities.
The loci where Higgs or mixed branches appear must have massless particles, but the appearance of new massless particles
does not necessary mean that there is a mixed branch. The method described in this paper detects mixed branches rather than massless particles. 
It would be very interesting to develop a systematic way to find massless particles and their types (e.g., mutually local or nonlocal, abelian or nonabelian, etc.).
The study of effective field theory  on the Higgs branch roots is particular interesting, as those singularities might 
survive under $\mathcal{N}=1$ deformation \cite{Argyres:1996eh}.
We leave the study of those effective theories and $\mathcal{N}=1$ deformation to the future.

\section*{Acknowledgments}
D.X would like to thank P. Argyres for helpful discussion. K.Y. would like to thank K. Maruyoshi for helpful discussion. 
D.X would like to thank KITP for hospitality during the final stage of this project. 
This research is supported in part by Zurich Financial services membership and by the U.S. Department of Energy, grant DE-SC0009988  (DX). 
The work of K.Y. is supported in part by NSF grant PHY- 0969448.

\appendix

\section{More on SQCD}\label{sec:appA}
In this appendix we study the $SU(N)$ SQCD with general numbers of flavors $N_f<2N$.

\subsection{Irregular singularities.}\label{sec:irregular}
We will need some irregular singularities, so we review them.

\paragraph{The rules.}
From a brane realization of $SU(N)$ SQCD with $K<N$ flavors~\cite{Witten:1997sc}, 
one can see that there exists an irregular singularity of the form~\cite{Gaiotto:2009hg}
\beq
\phi_i(z)=\frac{\Lambda^{N-K}}{ z^{1+N-K}} \delta_{i,N-K}+\CO(z^{-i}),\label{eq:simpleirreg}
\eeq
where $\Lambda$ is a parameter which is roughly a dynamical scale of the corresponding field theory.
This singularity has the flavor symmetry $U(K)$.

A slight generalization of this singularity is the following. We take a partition of $K$, $Y=[m_a]$ where $K=m_1+\cdots+m_\ell$.
Then we have a puncture labeled by $(K,Y)$.
Consider the dual partition $Y^D=[\tilde{m}_a]$. The singularity is now given by
\beq
\phi_i(z)=\frac{\Lambda^{N-K}}{ z^{1+N-K}} \delta_{i,N-K}+\CO(z^{-p_i}),\label{eq:irreg}
\eeq
where 
\beq
p_i=\left\{ \begin{array}{ll}
i, & ~~~2 \leq i \leq N-K, \\
i+1-a, &~~~(N-K)+\sum_{b=1}^{a-1}\tilde{m}_b < i \leq (N-K)+ \sum_{b=1}^{a}\tilde{m}_b 
\end{array}
\right.
\eeq
If $K=N-1$, we need to shift $x$ in the curve \eqref{eq:SWGcurve} so that the $x^{N-1}$ term is cancelled.

The rules discussed in section~\ref{sec:mix} is slightly changed as follows.
Take a partition $X=[n_s]$ used in the factorization of the curve \eqref{eq:facto}.
One of the numbers in $[n_s]$, say $n_r$,
must satisfy $n_r \geq N-K$. Then, we take a partition of $n_r-N+K$, denoted as $Y^D_r=[\tilde{m}_{r,a}]$ where $\sum_a \tilde{m}_{s,a}=n_r-N+K$,
and assume
\beq
\phi_{r,i} &=\frac{\Lambda^{N-K}}{ z^{1+N-K}} \delta_{i,N-K}+\CO(z^{-p_{r,i}}) \nonumber \\
p_{r,i} &=\left\{ \begin{array}{ll}
0 & ~~~i=1 \\
i & ~~~2 \leq i \leq N-K, \\
i+1-a &~~~(N-K)+\sum_{b=1}^{a-1}\tilde{m}_{r,b} < i \leq (N-K)+ \sum_{b=1}^{a}\tilde{m}_{r,b }
\end{array}
\right. . \label{eq:blockirre}
\eeq
The other blocks, i.e., $n_s$ with $s \neq r$, has a partition of $n_s$ denoted as $Y^D_s=[\tilde{m}_{s,a}]$, and the singularities
are given similar to \eqref{eq:degofsing} as in the regular puncture case. 
We combine the partitions $Y^D_s=[\tilde{m}_{r,a}]$ and $Y^D_r=[\tilde{m}_{s,a}]$ ($s\neq r)$
as discussed in section~\ref{sec:2} to get a partition of $K$ denoted as $Y'^D=[\tilde{m}'_a]$ where $\sum_{a} \tilde{m}'_a=K$.
Then we require that $Y'^D \leq Y^D$.

The local contribution to the quaternionic dimension of $\CH_\alpha$ is given by the formula
\beq
\dim_{\mathbb H} \CH_\alpha (p) = d(Y')-d(Y),\label{eq:localireghiggs}
\eeq
where $Y'$ is the partition of $K$ dual to $Y'^D$, and $d(Y)$ is defined as
\beq
d(Y) 
=\frac{1}{2} \left(K^2-\sum_{a=1}^{\ell} \tilde{m}_a^2 \right).
\eeq 
We stress that the relevant partitions are partitions of $K$ instead of $N$ in this irregular case.

\paragraph{Derivation.}
The above rules are derived as follows.
If we use complex gauge transformations $\Phi \to g \Phi g^{-1}$, $g \in SU(N)_{\mathbb C}$, to make the Higgs field of the Hitchin system $\Phi$ 
block diagonal, the irregular singularity is given as
(assuming that the singularity is at $z=0$),
\beq
 \Phi  \to &  \frac{1}{z} \left( \begin{array}{cc} 
 \nu_h & 0 \\
 0 & 0
 \end{array} \right)
+ \Lambda\left( \begin{array}{cc} 
0 & 0 \\
 0 & W(z)
 \end{array} \right) \label{eq:SUNirreg1}
\eeq
where $\nu_h$ is a $K \times K$ nilpotent matrix in the orbit $\nu_h \in \bar{O}_{Y^D}$,
and $W(z)$ is an $(N-K) \times (N-K)$ matrix whose eigenvalues are given as
\beq
W(z) \to \frac{1}{z^{1+1/(N-K)}}\diag(1,\omega_{N-K},\cdots, \omega_{N-K}^{N-K-1})+\cdots
\eeq
where $\omega_{N-K}=\exp ( 2\pi i/(N-K))$. The first $K \times K$ block behaves just as regular singularity.
The commutation relation $[\Phi,\vec{\varphi}]=0$ requires that the block which commutes with $W(z)$ must be proportional to 
an identity matrix. Along the lines of the discussion around \eqref{eq:blockirre}, this block of $\vec{\varphi}$ is an $n_r \times n_r$ 
unit matrix with $n_r \geq N-K$.

In \cite{Yonekura:2013mya}, it was proposed that the holomorphic moment map $\mu_h$ of the $U(K)$ flavor symmetry is related to $\varphi_h$ as
\beq
\det (x I_N- \varphi_h)  =
\det \left( x I_N -\left(
\begin{array}{cc}
\mu_h & 0 \\
0 & 0
\end{array}
\right)+ \frac{\tr \mu_h}{N} I_N \right). \label{eq:irregconst1}
\eeq
Also from this, we can see that $\phi_h$ must be proportional to the unit matrix in the block which contains $W(z)$.

Now the matrices $\nu_h$ and $\mu_h$ are just as in the regular case.
The rules stated above can be found as in the regular singularity.

\subsection{Mixed branches of SQCD}

Now we study mixed branches of SQCD using the irregular singularities discussed above.

\subsubsection{$N_f<N$}\label{eq:simpleSQCD}
Before going to the general cases, let us consider the case $N_f<N$ in a simple realization as a warm-up.
We proceed as if $N_f<N-1$, but the case $N_f=N-1$ can be obtained just by shifting $x$ to cancel the $x^{N-1}$ term in the curve.

On a Riemann sphere, we put the singularity \eqref{eq:simpleirreg} with $K=N_f$ at $z=0$, and with $K=0$ at $z =\infty$.
The curve is
\beq
0=x^N+\sum^{N}_{i=2}  \frac{u_i}{z^i}x^{N-i}+\frac{\Lambda^{N-N_f}}{z^{1+N-N_f}}x^{N_f}+\frac{\Lambda^N}{z^{N-1}}. \label{eq:nf<Ncurve}
\eeq
Because of the existence of the singularity at $z= \infty$ (i.e., the last term $\Lambda^N/z^{N-1}$ in the above equation), 
it is impossible to factorize the curve,  so we have $X=[n_s]=[N]$.
All the contributions to the Higgs factor comes from the puncture at $z=0$.

Let us apply \eqref{eq:blockirre} to the puncture at $z=0$ with $K=N_f$.
Because of the existence of the last term $\Lambda^N/z^{N-1}$ in \eqref{eq:nf<Ncurve} which has a pole of order $N-1$ at $z=0$, 
one can check that the partition must be of the form $Y'^D=[\tilde{m}'_a]=[N_f-r,r]$, where $0\leq r \leq [N_f/2]$. Then the curve is given as
\beq
0=x^N+\sum^{N-r}_{i=2}  \frac{u_i}{z^i}x^{N-i}+\frac{\Lambda^{N-N_f}}{z^{1+N-N_f}}x^{N_f}+\frac{\Lambda^N}{z^{N-1}}.
\eeq
The dimension of the Coulomb branch $\CC_r$ is $N-r-1$ spanned by $u_2,\cdots,u_{N-r}$.
The quaternionic dimension of the Higgs component $\CH_r$ is computed by using \eqref{eq:localireghiggs} as
\beq
\dim_{{\mathbb H}} \CH_r &= \frac{1}{2} \left( N_f^2-(N_f-r)^2-r^2 \right)=r(N_f-r).
\eeq
This is exactly as was found by Argyres, Plesser and Seiberg~\cite{Argyres:1996eh}.

\subsubsection{$N_f=N_1+N_2$}

Now let us consider more general cases.
We take $N_f=N_1+N_2$, and assume that $N_{1,2}<N$.

Using the singularity \eqref{eq:SUNirreg1} for $K=N_1,N_2$ at $z=0,\infty$ respectively, the curve is
\beq
0=x^N+\sum^{N}_{i=2}  \frac{u_i}{z^i}x^{N-i}+\frac{\Lambda^{N-N_1}}{z^{1+N-N_1}}x^{N_1}+\frac{\Lambda^{N-N_2}}{z^{-1+N-N_2}}x^{N_2}.
\eeq
When factorizing the curve as in \eqref{eq:facto}, there are two possibilities.
\begin{enumerate}
\item One factor contains both the higher singular terms at $z=0,\infty$, i.e., the factorization is of the form
\beq
0=\left( x^{n_1}+\frac{\Lambda^{N-N_1}}{z^{1+N-N_1}}x^{N_1-N+n_1}+\frac{\Lambda^{N-N_2}}{z^{-1+N-N_2}}x^{N_2-N+n_1} +\cdots \right)(\cdots).
\label{eq:mesonicfacto}
\eeq
\item Higher singular terms at $z=0$ and $z=\infty$ are contained in separate blocks,
\beq
0=\left( x^{n_1}+\frac{\Lambda^{N-N_1}}{z^{1+N-N_1}}x^{N_1-N+n_1}+\cdots \right)
\left( x^{n_2}+\frac{\Lambda^{N-N_2}}{z^{-1+N-N_2}}x^{N_2-N+n_2}+\cdots \right)(\cdots).\label{eq:baryonicfacto}
\eeq
This is possible only if $N_f=N_1+N_2 \geq N$.
\end{enumerate}
We call the first branch as mesonic branch and the second branch as baryonic branch.

\paragraph{Mesonic branch.}
Assume that the curve is factorized as in \eqref{eq:mesonicfacto}. Then the factors which do not contain the higher pole terms
have only regular singularities at $z=0$ and $\infty$. Then, one can see that the only possibility is the factorization of the form

\beq
0=x^{N-n_1}\left(x^{n_1}+\sum^{n_1}_{i=2} \frac{u_i }{z^i}x^{{n_1}-i}+
\frac{\Lambda^{N-N_1}}{z^{1+N-N_1}}x^{N_1-N+n_1}+\frac{\Lambda^{N-N_2}}{z^{-1+N-N_2}}x^{N_2-N+n_1}
\right).
\eeq
Then the partition specifying the factorization is given by $X=[n_s]=[n_1,1^{N-n_1}]$.

First, let us consider the case $n_1>\max \{N-N_1,N-N_2\}$. In this case, we can assume $u_{n_1} \neq 0$, since otherwise
the curve would be further factorized according to the partition $[n_s]=[n_1-1, 1^{N-n_1+1}]$  if $u_{n_1}=0$. 
Then, the condition $u_{n_1} \neq 0$ requires that the partitions $Y'^D=[\tilde{m}'_a]$ of $N_1$ and $N_2$ at $z=0$ and $z=\infty$
must be $[N_1-N+n_1,1^{N-n_1}]$ and $[N_2-N+n_1,1^{N-n_1}]$, respectively. Thus we get
\beq
\dim_{{\mathbb H}} \CH_\alpha=&(N- n_1)+ \frac{1}{2}\left( N_1^2-(N_1-N+n_1)^2-(N-n_1)\cdot 1^2 \right) \nonumber \\
&~~~+\frac{1}{2}\left( N_2^2-(N_2-N+n_1)^2-(N-n_1)\cdot 1^2 \right) \nonumber \\
=&r(N_f-r),
\eeq 
where we have defined $r=N-n_1$ and $N_f=N_1+N_2$. From the constraint $\max \{N-N_1,N-N_2\}<n_1 \leq N$,
we get $0 \leq r < \min\{N_1,N_2 \}$

Next, let us consider the case $n_1=\max \{N-N_1, N-N_2 \}$. Without loss of generality, we assume $N_1 \geq N_2$ and $n_1=N-N_2$.
The curve is 
\beq
0=x^{N_2}\left(x^{N-N_2}+\sum^{N-N_2}_{i=2} \frac{u_i }{z^i}x^{{N-N_2}-i}+
\frac{\Lambda^{N-N_1}}{z^{1+N-N_1}}x^{N_1-N_2}+\frac{\Lambda^{N-N_2}}{z^{-1+N-N_2}}
\right).
\eeq
Here we do not assume that $u_{N-N_2}$ is nonzero.

Now, the situation is somewhat similar to section~\ref{eq:simpleSQCD}.
The existence of the last term $\Lambda^{N-N_2}/z^{-1+N-N_2}$ makes it impossible to factorize the curve further.
Furthermore, because of this last term, the partition $Y'^D=[\tilde{m}'_a]$ of $N_1$ at $z=0$ 
must be of the form $[N_1-N_2-r',r',1^{N_2}]$, where $0 \leq r' \leq [(N_1-N_2)/2]$. The partition $Y'^D=[\tilde{m}'_a]$ of $N_2$ at $z =\infty $
is given by $[1^{N_2}]$. The nonzero Coulomb moduli are given by $u_2, \cdots, u_{N-N_2-r'}$, so the
dimension of the Coulomb factor $\CC_\alpha$ is $N-N_2-r'-1$. The dimension of the Higgs component $\CH_\alpha$ is
\beq
\dim_{{\mathbb H}} \CH_\alpha&=N_2+\frac{1}{2}\left(N_1^2- (N_1-N_2-r')^2-r'^2-N_2 \right)+\frac{1}{2}\left(N_2^2-N_2  \right) \nonumber \\
&=r(N_f-r),
\eeq
where we have defined $r=r'+N_2$. It satisfies $\min\{N_1,N_2\}=N_2 \leq r \leq [N_f/2]$.

Summarizing what we have found above, mesonic mixed Higgs-Coulomb branches are labelled by $r$, $0 \leq r \leq [N_f/2]$.
The Higgs branch dimension is $r(N_f-r)$, while the Coulomb branch dimension is $N-r-1$. This is exactly as was found in \cite{Argyres:1996eh}.

\paragraph{Baryonic branch.}
Next let us consider the case \eqref{eq:baryonicfacto}. In this case, one can check that the only possible curve allowed by the singularities is given by
\beq
0=x^{N_f-N}\left( x^{N-N_1}+\frac{\Lambda^{N-N_1}}{z^{1+N-N_1}} \right)
\left( x^{N-N_2}+\frac{\Lambda^{N-N_2}}{z^{-1+N-N_2}} \right).
\eeq
So all the Coulomb moduli are fixed. This factorization corresponds to the partition $X=[n_s]=[N-N_1,N-N_2, 1^{N_f-N}]$.

Expanding this curve, we find that $u_{2N-N_f}=\Lambda^{2N-N_f} \neq 0$. Then, 
the partitions $Y'^D=[\tilde{m}'_a]$ of $N_1$ and $N_2$ at $z=0$ and $z=\infty$ are given by
$[N-N_2, 1^{N_f-N}]$ and $[N-N_1,1^{N_f-N}]$, respectively. Therefore we get
\beq
\dim_{{\mathbb H}} \CH_\alpha &=(N_f-N+1)+\frac{1}{2}\left(N_1^2-(N-N_2)^2-(N_f-N) \right) \nonumber \\
&~~~~+\frac{1}{2}\left(N_2^2-(N-N_1)^2-(N_f-N) \right) \nonumber \\
&=N(N_f-N)+1.
\eeq
This reproduce the result of \cite{Argyres:1996eh}.

It is remarkable that the final results do not depend on $N_1$ and $N_2$ separately, but depend only on the combination $N_f=N_1+N_2$,
although each bulk and local contributions to $\dim_{\mathbb H} \CH_\alpha$ are different for different pairs $(N_1,N_2)$.
This means that different brane constructions lead to the same low energy 4d theory as expected.

There are also constructions of SQCD with $N_f \geq N$ by using two regular punctures (one of them is full and the other is simple),
and one irregular puncture. We leave that case for the reader.

\bibliographystyle{JHEP}
\bibliography{ref}

\end{document}